%% file: draft.tex
\documentclass[useAMS,usenatbib]{mn2e}

\usepackage[T1]{fontenc}
\usepackage{pslatex}
\usepackage{amsmath}
\usepackage{amsfonts}
\usepackage{color}
\usepackage{url}
\usepackage{graphicx}
\usepackage{subfigure}
\usepackage{amssymb}
\usepackage{textcomp}
\usepackage{soul}
\usepackage{booktabs}
\usepackage{array}
\usepackage{hyperref}
 \graphicspath{{images/}}

{\newif\ifnotend
\notendtrue
\def\veclist{ABCDEFGHIJKLMNOPQRSTUVWXYZabcdefghijklmnopqrstuvwxyz.}
\def\top#1#2.{#1}
\def\tail#1#2.{#2.}
\loop\expandafter\xdef\csname v\expandafter\top\veclist\endcsname%
{{\noexpand\bf\expandafter\top\veclist}}
\edef\veclist{\expandafter\tail\veclist}
\if\veclist.\notendfalse\fi\ifnotend\repeat}
\mathchardef\mhyphen="2D

\title[A model for periodic blazars]{A model for periodic blazars}
\author[Sobacchi, Sormani, \& Stamerra]{Emanuele Sobacchi$^1$, Mattia C. Sormani$^{2,3}$, Antonio Stamerra$^{1,4,5}$\\
$^1$Scuola Normale Superiore, Piazza dei Cavalieri 7, 56126 Pisa, Italy\\
$^2$Rudolf Peierls Centre for Theoretical Physics, 1 Keble Road, Oxford OX1 3NP, UK\\
$^3$Institute for Theoretical Astrophysics, Zentrum f\"ur Astronomie der Universit\"at Heidelberg, Albert-\"Uberle-Str. 2, 69120 Heidelberg, Germany\\
$^4$INAF, Osservatorio Astrofisico di Torino,  I-10025  Pino Torinese, Torino, Italy\\
$^5$ASI Science Data Center, via del Politecnico s.n.c., I-00133 Roma, Italy
}

\begin{document}

\date{}

\def\p{\partial}
\def\Omegap{\Omega_{\rm p}}

\newcommand{\di}{\mathrm{d}}
\newcommand{\bfx}{\mathbf{x}}
\newcommand{\bfe}{\mathbf{e}}
\newcommand{\vlos}{\mathrm{v}_{\rm los}}
\newcommand{\Tspin}{T_{\rm s}}
\newcommand{\Tb}{T_{\rm b}}
\newcommand{\degree}{\ensuremath{^\circ}}
\newcommand{\Th}{T_{\rm h}}
\newcommand{\Tc}{T_{\rm c}}
\newcommand{\bfr}{\mathbf{r}}
\newcommand{\bfv}{\mathbf{v}}
\newcommand{\pc}{\,{\rm pc}}
\newcommand{\kpc}{\,{\rm kpc}}
\newcommand{\Myr}{\,{\rm Myr}}
\newcommand{\Gyr}{\,{\rm Gyr}}
\newcommand{\kms}{\,{\rm km\, s^{-1}}}
\newcommand{\de}[2]{\frac{\partial #1}{\partial {#2}}}
\newcommand{\cs}{c_{\rm s}}
\newcommand{\rb}{r_{\rm b}}
\newcommand{\rqu}{r_{\rm q}}
\newcommand{\nuP}{\nu_{\rm P}}
\newcommand{\thetaobs}{\theta_{\rm obs}}
\newcommand{\hatn}{\hat{\textbf{n}}}

\maketitle

\begin{abstract}
We describe a scenario to explain blazar periodicities with timescales of $\sim$ few years. The scenario is based on a binary super-massive black hole (SMBH) system in which one of the two SMBH carries a jet. We discuss the various mechanisms that can cause the jet to precess and produce corkscrew patterns through space with a scale of $\sim$ few pc. It turns out that the dominant mechanism responsible for the precession is simply the imprint of the jet-carrying SMBH orbital speed on the jet. Gravitational deflection and Lense-Thirring precession (due to the gravitational field of the other SMBH) are second order effects.

We complement the scenario with a kinematical jet model which is inspired to the spine-sheath structure observed in M87. One of the main advantages of such a structure is that it allows the peak of the synchrotron emission to scale with frequency according to $\nu F\propto \nu^{\xi}$ as the viewing angle is changed, where $\xi$ is not necessarily 3 or 4 as in the case of jets with uniform velocity, but can be $\xi \sim 1$.

Finally, we apply the model to the source PG1553+113, which has been recently claimed to show a $T_{\rm obs}=(2.18\pm 0.08)\text{ yr}$ periodicity. We are able to reproduce the optical and gamma-ray light curves and multiple synchrotron spectra simultaneously. We also give estimates of the source mass and size. 
\end{abstract}

%%\gg%%%%%%%%%%%%%%%%%%%%%%%%%%%%%%%%%%%%%%%

\input{paper.tex}

\section*{Acknowledgements}

MCS deeply thanks Steve Shore for carefully reading several drafts of the manuscript and extensive comments and discussions. The authors thank Eugene Vasiliev, Fabrizio Tavecchio and Giacomo De Palma for useful discussions. The authors also thank Elisa Prandini for preliminary multifrequency results and Elina Lindfors and Kari Nillson for the publicly available optical data from the Tuorla blazar monitoring program.
This research has made use of the archives and services of the ASI Science Data Center (ASDC), a facility of the Italian Space Agency (ASI Headquarter, Rome, Italy). 
We thank the Swift team for making these observations possible, the duty scientists, and science planners. ES and MCS both acknowledge contributing equally to this work.

%%%%%%%%%%%%%%%%%%%%%%%%%%%%%%%%%%%%%%%%%
\def\aap{A\&A}\def\aj{AJ}\def\apj{ApJ}\def\mnras{MNRAS}\def\araa{ARA\&A}\def\aapr{Astronomy \&
 Astrophysics Review}\def\apjs{ApJS}\def\apjl{ApJ}\def\pasj{PASJ}\def\nat{Nature}\def\prd{Phys. Rev. D}
\def\ssr{Space Sci. Rev.}\def\pasp{PASP}
\bibliographystyle{mn2e}
\bibliography{bibliography}

\appendix
\section{Derivation of Eq.~\ref{EQF}}
\label{sec:appendix}

In this appendix we derive the flux observed in the frame $K$ (at rest with respect to the emission pattern $\Sigma$) by an observer and whose line of sight makes an angle $\thetaobs$ with the axis of the jet (Eq. \ref{EQF}).

As explained in the main text, we assume that all fluid elements emit with a given isotropic emissivity $j_0(\nu')$ in their rest frames. The emissivity in the frame $K$, in which the fluid is moving, can be obtained by applying the usual special relativistic transformation rules \cite[see for example][]{MihalasMihalas1984,rlightman} and is given by:
\begin{equation} j(\nu,\hatn) = \delta^2 j_0\left(\frac{\nu}{\delta}\right). \label{eq:j} \end{equation}
$\nu$ is the transformed frequency, i.e. the frequency in the frame $K$ of light whose frequency in the fluid element rest frame is $\nu'$. The relation between the two frequencies is
\begin{equation} \nu= \delta \nu'. \end{equation}
The rest of the notation is explained in the main text. To find the observed flux we need to sum contributions over all fluid elements. We assume that the jet is optically thin, thus the radiative transfer equation reduces to:
\begin{equation} I(\nu,\hatn) = \int j(\nu,\hatn) \di x,  \end{equation}
where $I(\nu,\hatn)$ is the specific intensity along the direction $\hatn$ in the frame $K$. The integral is performed along the line of sight in the direction $\hatn$, where $\di x$ is the length element along the line of sight. The final step to obtain the flux is to sum specific intensities from all line of sights intersecting the jet. For an observer located very far from the jet, at a distance $D\gg L$ where $L$ is the height of the cylinder, the flux is given by:
\begin{equation} F(\nu,\hatn) = \frac{1}{D^2} \int_{\Sigma}  j(\nu,\hatn) \di V \label{eq:F1} \end{equation}
where $\di V$ is the volume element and the integral is performed over the whole cylindrical emission pattern $\Sigma$. Note that this integral does not depend on the particular shape of the emission pattern, but only on its total volume and on the amount of material present at each velocity. Hence, our result is valid for other geometries provided that the total amount of material at velocity $\Gamma_1$ and $\Gamma_2$ are the same. 

Plugging Eq. \eqref{eq:j} into Eq. \eqref{eq:F1} and integrating over the volume of the cylindrical emission pattern we finally find
\begin{align} F(\nu,\hatn) =  \frac{\pi LR_2^2}{D^2} \left[ \lambda \delta_1^2 j_0\left(\frac{\nu}{\delta_1}\right) + (1-\lambda) \delta_2^2 j_0\left(\frac{\nu}{\delta_2}\right) \right] \end{align}
where $0\leq\lambda\leq1$ is given by:
\begin{equation} \lambda \equiv \left(\frac{R_1}{R_2}\right)^2\;. \end{equation}

\section{Data description and analysis} \label{data}

The {\it Swift} satellite \citep{Gehrels2004} observed PG~1553+113 since 2005 during outbursts and almost regularly since 2013. 15 observations from April 2013 to April 2016 have been chosen to cover the different phases of the periodic modulation. These are listed in Table \ref{tab:data}. The selected snapshots have been observed simultaneously with the the X-ray Telescope \citep[XRT;][0.2--10.0 keV]{Burrows2005}, and with all six filter of  the Ultraviolet/Optical Telescope \citep[UVOT;][170--600 nm]{Roming2005}.
The XRT data were processed using the FTOOLS task \texttt{xrtpipeline} (version 0.13.2), which is distributed by HEASARC within the HEASoft package (v6.19). Events with grades $0-12$ were selected for the data (see \citealt{Burrows2005}) and corresponding response matrices available in the {\it Swift} CALDB version were used.
The data were collected in photon counting mode (PC) and windowed timing mode (WT). 
When the source count rate in photon counting mode was higher than 0.6 counts s$^{-1}$ the  pile-up was evaluated following the standard procedure.\footnote{\url{http://www.swift.ac.uk/analysis/xrt/pileup.php}} Observations affected by pile-up were corrected masking the central region $7.1\text{ arcsec}$. The signal was extracted  within an annulus with inner radius of 3 pixels (7.1 arcsec) and outer radius of 30 pixels (70 arcsec). 
Events in different channels were grouped with the corresponding redistribution matrix (rmf), and ancillary (arf) files with the task \texttt{grppha}, setting a binning of at least 25 counts for each spectral channel in order to use the chi-squared statistics. The resulting spectra were analyzed with \texttt{Xspec} version 12.9.0n. We fitted the spectrum with an absorbed power-law using the photoelectric absorption model \texttt{tbabs}
\citep{Wilms2000}, with a neutral hydrogen column density fixed to its Galactic value \citep[$N_{\rm H} = 3.67\times10^{20}$\text{ cm}$^{-2}$;][]{HIdata}. 

UVOT data in the $v$, $b$, $u$, $w1$, $m2$, and $w2$ filters were reduced with the \texttt{HEAsoft} package v6.19 using the \texttt{uvotsource} task. We extracted the source counts from a circle with 5 arcsec radius centered on the source and the background counts from a circle with 30 arcsec radius in a near, source-free region. Conversion of magnitudes into dereddened flux densities was obtained by adopting the extinction value E(B--V) = 0.054 as in \citet{Raiteri:2015lr}, the mean galactic extinction curve in \citet{Fitzpatrick1999} and the magnitude-flux calibrations in \citet{Poole2008}.
Statistical uncertainty on magnitudes of the order of 0.03 mag, on the zero-point UVOT calibration $0.02-0.06\text{ mag}$ and the count rate to flux correction \citep[see for example][]{Poole2008} have been propagated to estimate the error on the flux, resulting in a 4\% to 6\% uncertainty.

\begin{table}
\caption{\small Summary of the observations performed with {\em Swift} on PG~1553+113 and used in this work. Obs ID: unique {\em Swift} ID identifying the pointing. MJD: Modified Julian date. Duration of the observation, in ksec. XRT observation mode; PC: photon counting, WT: windowed timing.}
\begin{tabular}{c c c c }
\toprule
Obs ID & MJD & Duration (ks) & XRT Mode\\
\midrule
00031368051 & 56393 & 2.2 & PC \\
00031368053 & 56420 & 2.0 & PC \\
00031368061 & 56455 & 2.0 & WT \\
00031368069 & 56749 & 1.1 & WT \\
00031368076 & 57022 & 2.1 & WT \\
00031368081 & 57032 & 2.0 & WT \\
00031368091 & 57091 & 0.8 & WT \\
00031368092 & 57094 & 1.2 & WT \\
00031368100 & 57109 & 1.4 & WT \\
00031368104 & 57147 & 1.2 & PC \\ 
00031368111 & 57200 & 1.0 & PC \\
00031368126 & 57261 & 1.7 & PC \\
00031368130 & 57301 & 1.7 & PC \\
00031368132 & 57391 & 1.4 & PC \\
00031368141 & 57471 & 1.4 & PC \\
\bottomrule
\end{tabular} \label{tab:data}
\end{table}

\end{document}

%% file: paper.tex
\section{Introduction} \label{sec:introduction}

Astrophysical jets are common and ubiquitous: they are found in systems with extremely different spatial scales, from kpc-long jets in Active Galactic Nuclei (AGN) to smaller jets in stellar systems such as x-ray binaries and protostars \cite[see for example][]{Livio2009}. 

In some cases jets are observed to precess, forming a corkscrew pattern through space. The most spectacular example is probably the microquasar SS 433 \citep{Margon1984,BlundellBowler2004,BlundellBowler2005}. Jets powered by AGNs can also show similar corkscrew patterns: an example is 3C 273 \citep{Bahcall++1995}, the first quasar ever discovered \citep{Hazard++1963,Schmidt1963,Kellermann2014,Hazard++2015}.
% although in this case the helical pattern is more commonly interpreted as due to Kelvin-Helmholtz instability rather than to jet precession \citep{Bahcall++1995,Perucho++2006}. 

Blazars are powerful gamma-ray sources that are believed to consist in a jet powered by an AGN pointing in the general direction of the Earth \citep[see for example][]{UrryPadovani1995, Marscher2013}. It is natural to ask whether their jets can precess too and, if so, what is the physical mechanism responsible for the precession. The chief observational signature of such precession would be a periodic signal, so the first step is to ask whether such periodicities are observed.

Blazars observed fluxes display extreme variability at all wavelenghts on a wide range of timescales, which are generally divided into three classes: microvariabilities (timescale of less than a day), short-term variabilities (timescale from days to few months) and long-term variabilities (timescale $\sim$ few years) \citep[e.g.][]{Urry1996, Gupta2014}. Information on longer-term variabilities is obviously not observationally accessible (yet) as we have been able to observe these objects for a few decades only.

Signs of periodicity are often claimed on all possible timescales \citep[e.g.][]{Urry2011,Gupta2014}. The question is usually whether such periodicities are genuine or mere statistical fluctuations. Very few examples show convincing periodicities to date. In this work we focus on long-term periodicities with timescales of $\sim$ few years. One of the most convincing case is that of OJ 287, which shows a periodic flaring on a timescale of $\sim$ 12 years. What makes this case particularly convincing is that the next outburst was predicted on the basis of the light curve of the preceding decades by \cite{Sillanpaa++1988}, and then actually observed by \cite{Sillanpaa++1996}. Other notable cases are
% PG 1302-102 \citep{Graham++2015,DOrazio++2015},
PKS 2155-304 \citep{Zhang++2014,Sandrinelli++2016}, 3C 279 \citep{Li++2009,Sandrinelli++2016} and PG 1553+113 \citep{Ackermann++2015}. Thus, there is promising indication that a genuine periodicity can be detected and it is worth investigating the problem from the theoretical point of view. 

In this paper we describe a scenario in which a binary SMBH can generate an observable jet periodic precession. In this scenario, the jet is ballistic and the angle between the jet axis and the observer's line of sight changes periodically with time. We discuss the various mechanism that can cause the angle to vary. The presence of a binary SMBH system is expected in a significant fraction of AGN \cite[e.g.][]{Begelman++80}, and a binary system has been often indicated as a possible source of year-like periodicities \cite[e.g.][]{Gupta2014,Graham++2015}. 

We complement our scenario with a jet model that has a spine-sheath structure and we discuss a phenomenological model for the jet synchrotron emission. The spine-sheath structure is inspired to M87 where a similar structure has been actually observed (\citealt{Kovalev++2007}; see also \citealt{TavecchioGhisellini2008}). However, nothing prevents the scenario to being complemented by models involving different jet structures and/or emission models.

In the second part of the paper we apply our scenario to the blazar PG 1553+113, who has recently been found to show a quasi-periodic behavior \citep{Ackermann++2015}, most prominently in the gamma-ray band. We focus on this source because our work originally developed as an attempt to understand it. Our scenario has however more general validity and we point to further candidates for its application.

The paper is organized as follows. In Section \ref{sec:jetmodel} we describe our simple spine-sheath jet model. We derive formulas for the predicted flux and we model the synchrotron emission phenomenologically. The scenario and the physical mechanisms responsible for the jet precession are explained in Section \ref{sec:binary}. We also discuss the predicted light curves and give formulas to estimate the model parameters. In Section \ref{sec:application} we apply our model to the blazar PG 1553+113. Finally, in Section \ref{sec:conclusion} we draw our conclusions and discuss potential applications and future work.

\section{Jet Model} \label{sec:jetmodel}

In this section we describe our jet model.\footnote{We discuss only emission from the jet and we do not consider any emission from the accretion disk and/or the host galaxy, which are largely subdominant with respect to the beamed radiation in blazar jets.} %The scenario that explains why the axis of the jet can rotate as seen from an inertial observer located at infinity is not discussed in this section, but in Section \ref{sec:binary} below.
The model is purely kinematical and we start by describing its geometrical structure. Then we consider the emissivity of material inside the jet and perform some simple calculations to derive the flux received by an observer whose line of sight makes an angle $\theta_{\rm obs}$ with the axis of the jet. %In this first part, we derive all equations as a function of an unknown rest-frame emissivity $j_0$, which is assumed to be given.
In the last part we study in more detail the synchrotron emission. We derive a phenomenological formula for the emissivity and discuss some characteristics of the spectrum predicted by this formula.

For simplicity we discuss the model by considering the case of a nearby source (i.e. at redshift $z\sim 0$). In applying it to real objects, one has to correct the relevant equations to take into account their actual redshift (see the discussion in Section \ref{sec:application}).

\subsection{Geometry of the jet}

\begin{figure}
\centering
\def\svgwidth{\columnwidth}
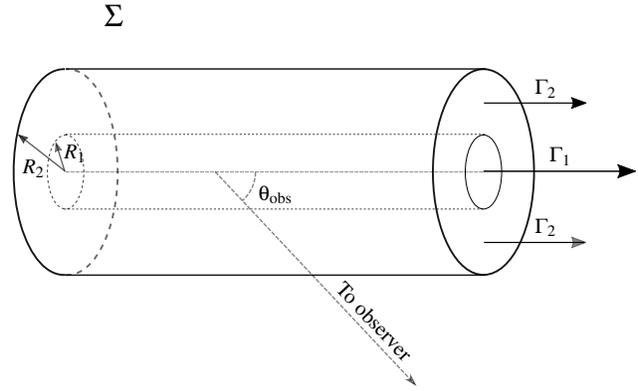
\caption{Schematic illustration of the jet structure.}
\label{fig:jet}
\end{figure}

We assume that material inside a cylindrical region $\Sigma$ (``emission pattern'') emits, while material outside it does not. The emission pattern is assumed to be stationary in some inertial frame $K$ but material is assumed to be steadily moving in the same frame (see Fig. \ref{fig:jet}) \citep[see also Section II of][]{LindBlandford1985}.\footnote{Note that the observed flux is different in the two following apparently equivalent situations: (i) the emission pattern is stationary and the fluid inside moves with uniform velocity $\bfv$ (i.e., it turns ``on and off'' as it enters and leaves the emission pattern) (this is the case we are discussing in this paper); (ii) the emission pattern moves together with the fluid at the same velocity $\bfv$. The rest-frame emissivity and all the other characteristics of the fluid are assumed to be exactly the same in both cases. The fact that the observed flux differs can be counterintuitive given that the two situations can be instantaneously identical (i.e., the emitting fluid instantaneously occupies the same spatial region and has at each point the same velocity and the same emissivity in both cases). The reason for the difference ultimately lies in the fact that at any given time an observer sees light emitted at various different previous times from different parts of the source, so how the whole pattern is moving does matter. See for example \cite{LindBlandford1985}.\label{footnote:1}} In this section the frame $K$ will also be identified with the observer's frame. 

The velocity of the fluid is assumed to be parallel to the axis of the cylinder. We assume a ``spine-sheath'' structure. This means that the jet has a ``spine'' (the central part, closer to the axis of the cylinder) where the fluid moves faster and a ``sheath'' (the external part) where the fluid moves slower. We assume that the Lorentz factors of material inside the jet are:

\begin{equation}
\begin{cases}
\Gamma_1 \quad {\rm if} \quad r \leq R_1, \\
\Gamma_2 \quad {\rm if} \quad R_1 \leq r \leq R_2,
\end{cases}
\end{equation}
where
\begin{equation}
\Gamma_{\rm i} = \frac{1}{\sqrt{1 - \beta_{\rm i}^2}}, \qquad i = 1,2
\end{equation}
where $\beta_{\rm i}$ is the velocity of the fluid in units of the speed of light. In our model, we assume that $\Gamma_1 \gg  \Gamma_2$. The idea is that in the sheath the fluid is slowed down by viscosity and/or by some other process. However, our model is purely kinematical and we do not discuss the hydrodynamics involved. The spine-sheath structure has been studied theoretically by a number of authors \citep{Sol++1989,Celotti++01,Ghisellini++05,Sikora++2016} and there is evidence that this structure does indeed occur in observed jets, for example the one of M87 (\citealt{Kovalev++2007}; see also \citealt{TavecchioGhisellini2008}).

\subsection{Emission from the jet}

We assume that all fluid elements emit with a given isotropic emissivity $j_0(\nu')$ in their rest frames. For the moment, we do not specify the form of $j_0(\nu')$. The emissivity in the frame $K$, where the fluid is moving, can be obtained by applying the usual special relativistic transformations, and will no longer be isotropic due to relativistic beaming.

Assuming an optically thin jet, an observer located very far from the jet ($D\gg L$ where $D$ is the distance of the observer and $L$ is the height of the cylinder) in the frame $K$ will then observe a flux given by (see Appendix \ref{sec:appendix} for a detailed derivation)
\begin{align} F(\nu,\hatn) =  \frac{\pi LR_2^2}{D^2} \left[ \lambda \delta_1^2 j_0\left(\frac{\nu}{\delta_1}\right) + (1-\lambda) \delta_2^2 j_0\left(\frac{\nu}{\delta_2}\right) \right] \label{EQF} \end{align}
where 
\begin{itemize}
\item $\hatn$ is the unit vector in the direction in which we are observing the emission in the frame $K$.
\item $\nu$ is the frequency in the frame $K$. 
\item $\delta_{\rm i}$ ($i=1,2$) is the Doppler factor, given by
\begin{equation} \delta_{\rm i}=\frac{1}{\Gamma\left(1-\hat{\textbf{n}}\cdot\pmb{\beta_{\rm i}}\right)},  \end{equation}
where
\begin{equation} \Gamma_{\rm i} = \frac{1}{\sqrt{1- \pmb{\beta_{\rm i}}^2 }}. \end{equation}
\item $0\leq\lambda\leq1$ is a weight which tells the relative contribution of the spine and the sheath
\begin{equation} \lambda \equiv \left(\frac{R_1}{R_2}\right)^2\;. \end{equation}
When $\lambda=0$ there is only sheath, when $\lambda=1$ there is only spine. The first term between square brackets in Eq. \eqref{EQF} originates from the spine, while the second term originates from the sheath.
\end{itemize} 

We can integrate Eq. \eqref{EQF} over a frequency band $\mathcal{B}=[\nu_{\rm a},\nu_{\rm b}]$ to obtain:
\begin{equation}
 F_{\mathcal{B}} = \frac{\pi L R_2^2}{D^2} \left[ \lambda \delta_1 ^3  \left( \int_{\nu_{\rm a}/\delta_1}^{\nu_{\rm b}/\delta_1} j_0 \left( y \right) \di y \right) + (1- \lambda) \delta_2 ^3  \left( \int_{\nu_{\rm a}/\delta_2}^{\nu_{\rm b}/\delta_2} j_0 \left( y \right) \di y \right) \right]\label{eq:FB1} \end{equation}
How does the flux predicted by Eq. \eqref{eq:FB1} change if we change the viewing angle $\thetaobs$? Note that the dependence of the observed flux on the viewing angle $\thetaobs$ comes entirely from the $\delta$'s. Taking $\pmb{\beta_{\rm i}} = (\beta_{\rm i}, 0, 0)$ and $\hatn = (\cos\thetaobs, \sin\thetaobs,0)$ these can be rewritten as
\begin{equation}
\delta_{\rm i} = \frac{1}{\Gamma_{\rm i} (1 - \beta_{\rm i} \cos\thetaobs)}, \quad i=1,2.
\label{eq:delta}
\end{equation}
%
%Hence the form of the dependence on $\thetaobs$ depends on the integrals between square parentheses in Eq. \eqref{eq:FB1}.

Let us consider a few examples. If we look at the total flux integrated over all possible frequencies, we have
\begin{equation}
 F_{\rm tot} = \frac{\pi L R_2^2}{D^2}  \left( \int_{-\infty}^{+\infty} j_0 \left( y \right) \di y \right) \left[ \lambda \delta_1 ^3  + (1- \lambda) \delta_2 ^3 \right] \label{eq:FB2} \end{equation}
In this case the dependence on $\thetaobs$ does not depend on the form of $j_0$ and is entirely contained in the $\delta$ factors inside the square parentheses. As a second example, consider a band over a limited frequency range and a power law emissivity $j_0 \propto K \nu^{\eta-1}$ where $\eta$ is a constant. In this case we obtain
\begin{equation}
 F_{\mathcal{B}} \propto \left[ \lambda \delta_1 ^{3-\eta}  + (1- \lambda) \delta_2 ^{3-\eta} \right]\;. \label{eq:FB3} \end{equation}
In this case the dependence on $\thetaobs$ does depend on the form of $j_0$ through $\eta$.

\subsection{Synchrotron emission}

\subsubsection{Phenomenological emissivity}

The emissivity $j_0(\nu')$ for the synchrotron emission can be derived phenomenologically in the following way. \cite{Massaro++2004} have shown that the observed spectrum of many BL-Lac objects is phenomenologically well described by a log-parabolic distribution:
\begin{equation}
\label{eq:sed}
\log\left[\nu F\left(\nu\right)\right]=\log\left[\nu_{\rm P}F\left(\nu_{\rm P}\right)\right]-b\left[\log\left(\frac{\nu}{\nu_{\rm P}}\right)\right]^2\;,
\end{equation}
where $\nu_{\rm P}$, $\nu_P F\left(\nu_{\rm P}\right)$ and $b$ are free parameters. To derive $j_0$, we assume that when $\Gamma_1=\Gamma_2=1$ the Eq. \eqref{EQF} and \eqref{eq:sed} give the same result. We obtain
\begin{equation}\label{eq:j0} 
j_0 \left(\nu'\right) = \frac{\nu_{\rm P} F(\nu_{\rm P}) D^2  }{ \pi \nu' L R_2^2} \exp\left\{ - b \left[\log\left(\nu'/\nu_{\rm P}\right) \right]^2 \right\}\;.
\end{equation}
We can check a posteriori that when this form of $j_0$ is used in Eq. \eqref{EQF} it produces fluxes that are very similar to the phenomenological Eq. \eqref{eq:sed}. 

\subsubsection{Spectra} \label{sec:flux}

\begin{figure}{\vspace{3mm}} 
\centering
\includegraphics[width=0.5\textwidth]{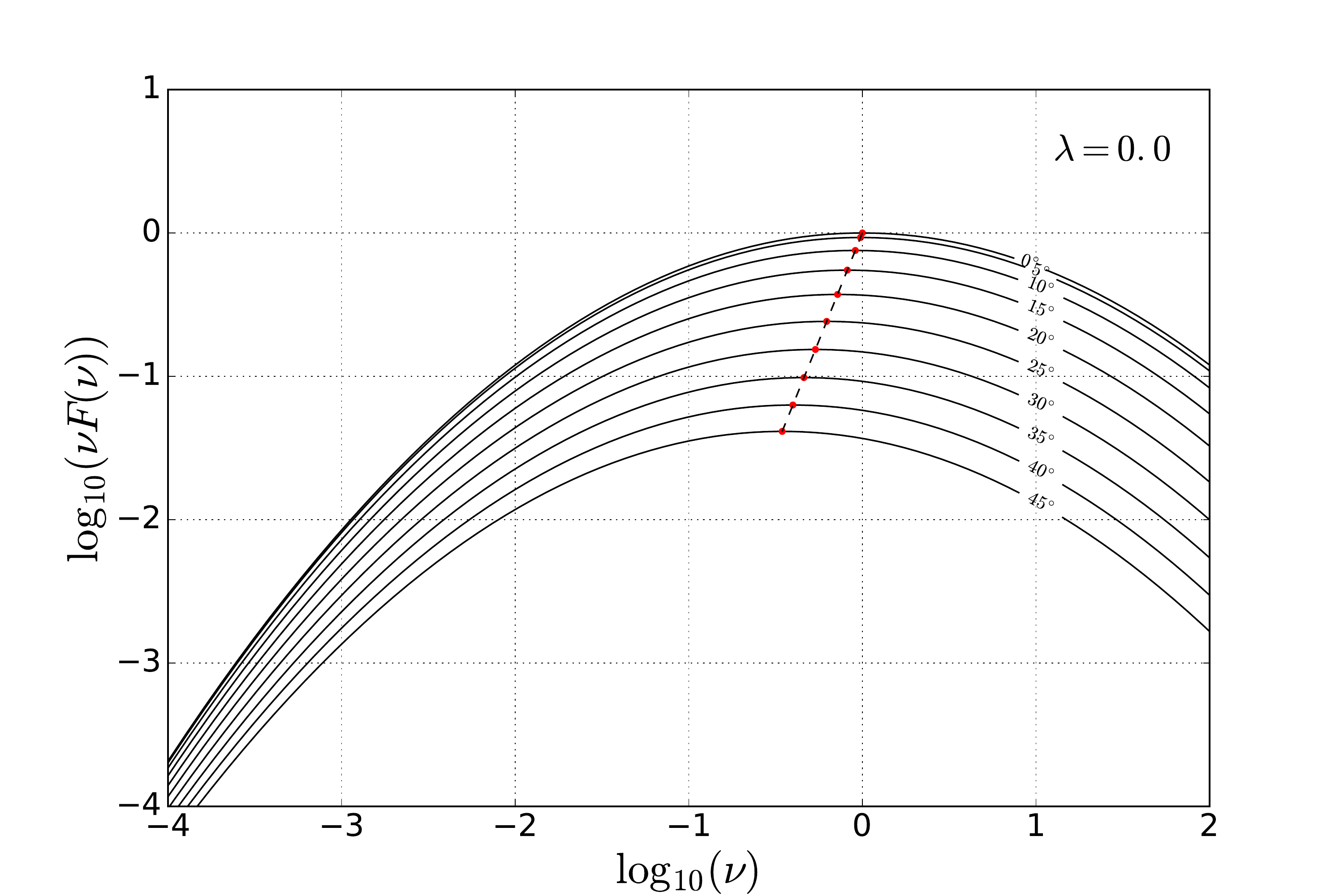}
\includegraphics[width=0.5\textwidth]{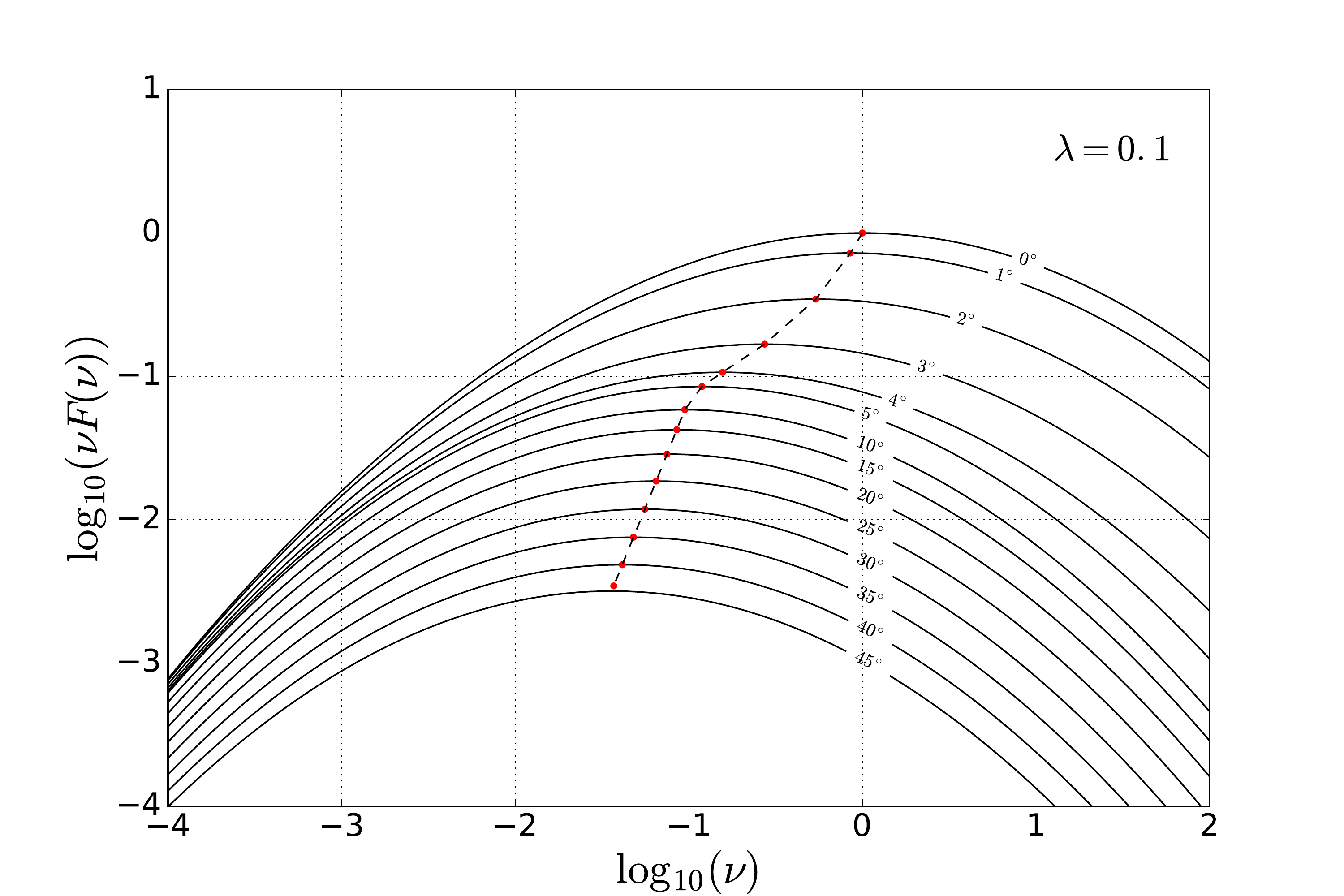}
\includegraphics[width=0.5\textwidth]{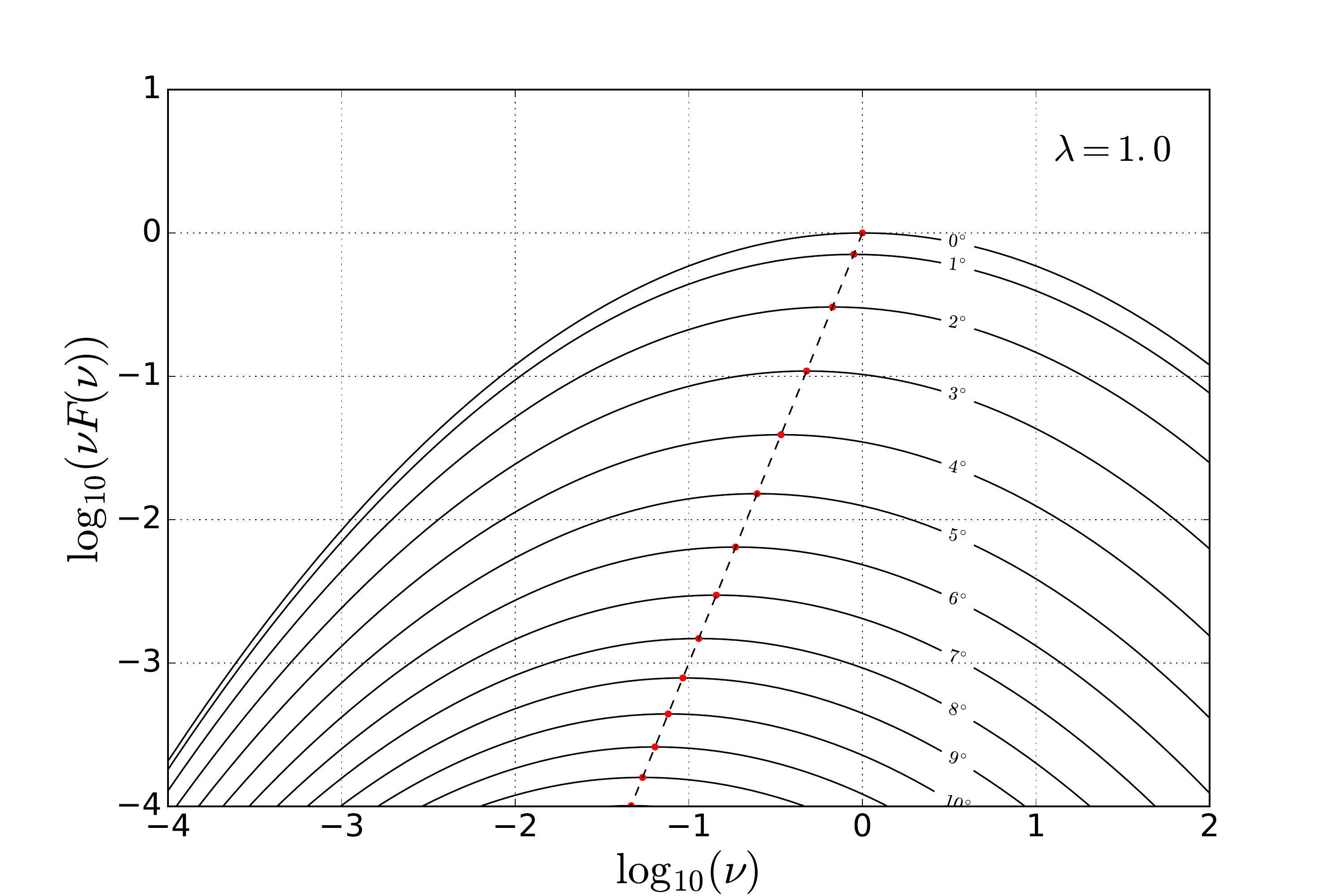}
\caption{Synchrotron spectra predicted by our jet model, calculated from Eq. \eqref{eq:F3}. The only difference between different panels is the value of $\lambda$. Each solid line is a spectrum obtained for a given value of $\thetaobs$. The maximum flux is obtained for $\thetaobs=0 \degree$ and other spectra are obtained for increasing values of $\thetaobs$ as shown in the figure. The red dots mark the peaks of the spectra. The black dashed line simply connects the red dots for easier visualization. The value of the other parameters are as follows: $\Gamma_1=20$, $\Gamma_2=2$, $b=1.0$. The curves are normalized so that the peak for $\thetaobs=0$ has a value of $\nu F(\nu)=1$ and occurs at a value $\nu=1$.}
\label{fig:predictedflux}
\end{figure}

Substituting Eq. \eqref{eq:j0} into Eq. \eqref{EQF} we obtain the flux of the synchrotron emission
\begin{align} \nonumber \frac{\nu F(\nu,\hatn)}{\nu_{\rm P} F(\nu_{\rm P})}  	& = \lambda \delta_1^3  \exp\left\{ - b \left[\log\left(\frac{\nu}{\delta_1 \nu_{\rm P}}\right) \right]^2 \right\} \\ 
						& + \left(1- \lambda\right) \delta_2^3  \exp\left\{ - b \left[\log\left(\frac{\nu}{\delta_2 \nu_{\rm P}}\right) \right]^2 \right\} .\label{eq:F3} \end{align}
The free parameters for the synchrotron emission in our model are:
\begin{itemize}
\item $\Gamma_1$, $\Gamma_2$: the speed of the inner and outer parts.
\item $\thetaobs$: the viewing angle.
\item $\nu_{\rm P} F(\nu_{\rm P})$: an overall normalization factor.
\item $\nu_P$: the peak frequency of the emissivity in the rest frame of the emitting material.
\item $b$: the curvature of the $\nu F(\nu)$ curve.
\end{itemize}

In Fig. \ref{fig:predictedflux} we show examples of spectra predicted by Eq. \eqref{eq:F3}. Each solid line is the spectrum predicted for a given value of the parameters listed above, and all such lines resemble very well a log-parabolic distribution, Eq. \eqref{eq:sed}. This confirms that our phenomenological $j_0$ does reproduce sensible spectra. The parameters used are listed in the figure caption.

Panels in Fig. \ref{fig:predictedflux} show how the observed emission changes as we change the viewing angle $\thetaobs$. Even if in its rest frame the material emits isotropically, in a frame where it moves with a relativistic factor $\Gamma$ its emission will be strongly beamed into a cone of angle $\simeq 1 / \Gamma$.

The top and bottom panels shows the case $\lambda=0$ and $\lambda=1$ respectively. Here either only the spine or sheath is present and all material moves with a single Lorentz factor. The maximum flux is obtained for $\thetaobs=0 \degree$, when we are looking at the jet right down its axis. As $\thetaobs$ is increased, the spectrum moves down and left in the $\log(\nu F) - \log(\nu)$ plane. The red dots trace how the peak of the spectrum moves. These dots form a straight line with a slope of $\xi=3$ (this can also be easily inferred directly from Eq. \ref{eq:F3}).\footnote{Note that this result is slightly different from the slope $\xi=4$ that would be obtained if the radiation were produced by a moving bubble whose emission pattern is moving together with the fluid instead of being stationary in the frame $K$ (see case ii in footnote \ref{footnote:1}). The latter is also a commonly discussed model for jets in blazars \citep[see for example][]{Ghisellini++98}.} The only difference between $\lambda=1$ and $\lambda=0$ is that in the first case the spectrum moves much more rapidly as a function of $\thetaobs$: the spine has a much higher factor than the sheath ($\Gamma_1 = 20 \gg \Gamma_2 = 2$) and thus relativistic beaming is much stronger when $\lambda=1$.

The middle panel shows the true spine-sheath case, where material at both $\Gamma_1$ and $\Gamma_2$ is present. The observed emission is now a combination of the emission from the spine and sheath regions, which are moving with different speeds. When $\thetaobs=0 \degree$, the spine dominates the emission thanks to its stronger beaming ($\lambda \delta_1^3 \simeq 6000 \gg (1-\lambda)\delta_2^3 \simeq 50$). When $\thetaobs$ is increased, the emission from the spine decreases quickly, while the emission from the sheath decreases much more slowly. Emission from the two regions combine to give a slope for the line of peaks (red dots) of $\xi \simeq 1$. This lasts until the emission from the spine becomes negligible, the line of peaks forms a knee and starts again sloping steeply with $\xi \simeq 3$ as in the case when only the sheath is present. The transition happens around $\thetaobs = 7\degree$ for this particular choice of the parameters.

How do the other parameters affect the curves in Fig. \ref{fig:predictedflux}? The parameters $\nuP$ and $F(\nuP)$ simply shift the curve in the $\log(\nu F) - \log(\nu)$ plane. The parameter $b$ changes the curvature of the log-parabolic shapes, but affects very weakly other characteristics of the spectra.

\section{Scenario for jet apparent precession} \label{sec:binary}

In this section we consider a scenario involving a binary SMBH system, in which one of the two SMBH carries a jet. We discuss the various physical mechanism responsible for its precession (on a $\sim$ few years timescale) and the creation of helical structures (on a $\sim$ few pc scale). We discuss the observational consequences on the spectrum and on the light curve.

\subsection{Hypotheses}

\begin{figure}
\centering
\def\svgwidth{\columnwidth}
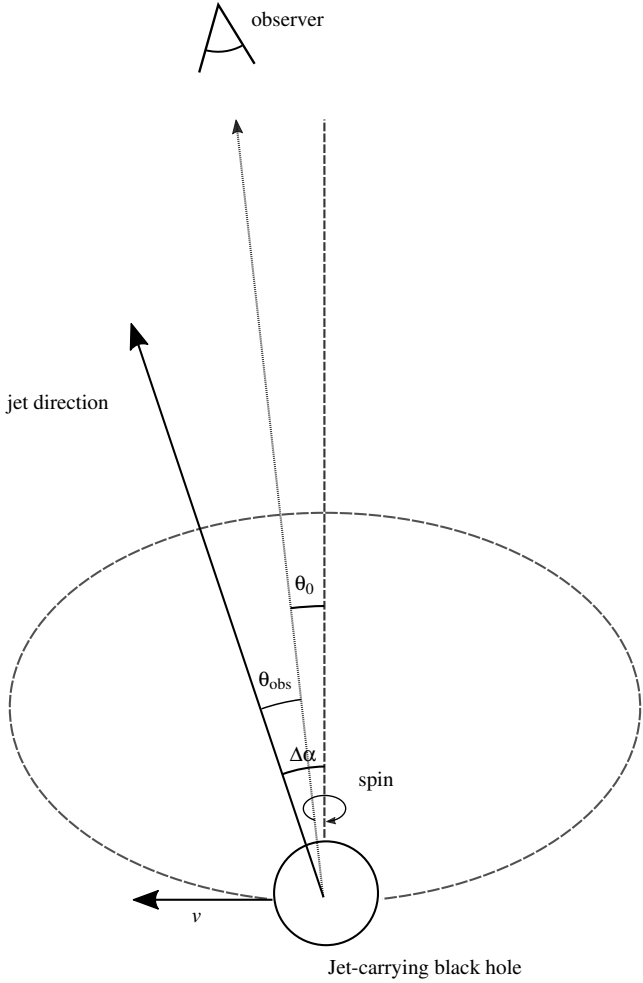
\caption{Schematic representation of the SMBH and its jet. The long-dashed line is the circular orbit followed by the black hole. The spin of the black hole is perpendicular to the orbital plane. The jet direction in the frame of the center of mass of the binary system makes an angle $\Delta \alpha$ with the orbital angular momentum.}
\label{fig:motion}
\end{figure}

Consider a binary system of SMBHs on circular orbits. For simplicity we assume that both the SMBHs have a spin that is perpendicular to the plane of the orbit. We assume that one of the SMBH carries a jet, and that the material is ejected in the direction of the spin in the SMBH rest frame (see Fig. \ref{fig:motion}). The jet is assumed to be purely ballistic.

The parameters describing this system are:
\begin{itemize}
\item $M$: the total mass of the system.
\item $q$: the mass ratio. The mass of the SMBH carrying the jet is $M_{\rm A} = M/\left(1+q\right)$, the mass of the other is $M_{\rm B} = M q /\left(1+q\right)$.
\item $R$: the separation between the SMBH, assumed to be larger than a few Schwarzschild radii.
\end{itemize}
A purely Newtonian calculation gives for the orbital period of the system:
\begin{equation}
\label{eq:period}
T=2\pi\left(\frac{R^3}{GM}\right)^{1/2}\;,
\end{equation}
where $G$ is the gravitational constant.

\subsection{Jet deviation and rotation}

Even if the jet plasma is emitted along the direction of the spin (i.e., perpendicularly to the orbital plane) in the rest frame of the jet-carrying SMBH, the jet will not be perpendicular to the orbital plane according to an infinitely distant observer due to a number of effects. 

\subsubsection{Dominant effect}

The dominant effect (i.e., the one that causes the jet angle to vary by the greatest amount) is simply the imprint of the SMBH orbital velocity on the jet: since the jet-carrying black hole is moving (on a circular orbit) with velocity $v$, the highly-relativistic ejected material will also possess the same velocity component in the observer's rest frame. The orbital velocity of the emitting SMBH is:
\begin{equation} \label{eq:v_SMBH}
v=\frac{q}{1+q}\left(\frac{GM}{R}\right)^{1/2}
\end{equation}
Hence, if we assume that the material is highly relativistic and hence is emitted at a velocity very close to that of light, the direction of the ejected plasma forms a small angle
\begin{equation}
\Delta\alpha \simeq \frac{v}{c}=\frac{q}{1+q}\left(\frac{GM}{Rc^2}\right)^{1/2}
\label{eq:alpha}
\end{equation}
with respect to the orbital angular momentum (see Fig. \ref{fig:motion}). During the orbital motion the direction of the ejecta slides on the surface of a cone with half-opening angle $\Delta \alpha$. Hence, the jet can rotate with respect to a distant observer, depending on its position. Assuming that the line of sight forms an angle $\theta_0\gtrsim\Delta\alpha$ with respect to the orbital angular momentum, the region at the base of the jet is observed at an angle $\thetaobs$ oscillating with an amplitude
\begin{equation}
\label{eq:deltatheta}
\Delta\thetaobs=2\Delta\alpha=\frac{2q}{1+q}\left(\frac{GM}{Rc^2}\right)^{1/2}
\end{equation}
and the same period $T$ of the orbital motion.\footnote{Note that if the line of sight is exactly parallel to the orbital angular momentum, the distant observer will see no rotation, i.e. the line of sight makes a constant angle with the jet (see also Eq. \ref{eq:costhetaobs}). In general, if $\theta_0\lesssim\Delta\alpha$, $\thetaobs$ oscillates with an amplitude $\Delta\thetaobs=2\theta_0$; thus one eventually finds $\Delta\thetaobs=2\min(\Delta\alpha,\theta_0)$.}

\subsubsection{General relativistic effects} \label{sec:GR}

The presence of a SMBH companion deflects the trajectory of the relativistic ejecta for the same reason light is deflected by a gravitational field. Since motion in a gravitational field depends only on the initial speed and position and not on the mass, a highly relativistic particle whose speed is very close to $c$ will follow the same path as a light ray. Hence at first order the deflection angle can be calculated following the common treatment for weak gravitational lensing:
\begin{equation}
\Delta\alpha_1=\frac{2}{c^2}\int_0^{+\infty}\frac{\partial}{\partial R}\phi\left(R,z\right)\text{d}z=\frac{2q}{1+q}\frac{GM}{Rc^2}\;,
\end{equation}
where we have assumed that the particle is launched from the position of the jet-carrying SMBH perpendicularly to the orbital plane and
\begin{equation}
\phi\left(R,z\right)=-\frac{q}{1+q}\frac{GM}{\sqrt{R^2+z^2}}
\end{equation}
is the gravitational potential of the mass $qM/\left(1+q\right)$ which is deflecting the ejecta.\footnote{We have verified that the self-deflection due to the gravitational field of the SMBH carrying the jet is smaller by a factor $\left(R_{\rm s}/R\right)\ln\left(R/R_{\rm s}\right)$ compared to the deflection from the other SMBH, where $R_{\rm s}$ is the Schwarzschild radius of the SMBH carrying the jet.} It is easy to verify that $\Delta\alpha_1\ll\Delta\alpha$ if the two SMBH are separated by more than a few Schwarzschild radii, a condition which can be safely assumed to be verified.

We have assumed that the spin of the individual SMBHs are parallel to the orbital angular momentum. However if we relax this assumption and take the spin of the jet-carrying SMBH not perpendicular to the orbital angular momentum, then its direction (and consequently the direction of the ejecta) can precede around the orbital angular momentum due to other effects. One of these is the Lense-Thirring effect (\citealt{Thirring1918,LenseThirring1918,Mashhoon++1984}; see also \citealt{MTW1973}) caused by the gravitational field of the companion SMBH. Taking the average angular velocity for the precession, $\Omega_{\rm LT}$, from \citet{BarkerOconnell75}, we calculate the angular deviation of the spin after one period as
\begin{equation}
\Delta\alpha_2\lesssim\Omega_{\rm LT}T=\frac{\pi q\left(3q+4\right)}{\left(1+q\right)^2}\frac{GM}{Rc^2}\;.
\end{equation}
We have $\Delta\alpha_2\ll\Delta\alpha$ if the two SMBHs are separated by more than a few Schwarzschild radii. Thus, over the time of a single orbit the deflection due to the imprint of the orbital speed is much greater than the deflection due to the Lense-Thirring effect. However, deflections due to LT can sum up over many periods to produce a total deviation equal to the misalignment between the spin and the orbital angular momentum. The timescale for this is
\begin{equation}
T_{\rm LT} = \frac{2 \pi}{\Omega_{\rm LT}} = \frac{2 (1+q)^2}{q (3q+4)} \frac{T}{GM/R c^2} \simeq 10^3\frac{q}{3q+4}\left(\Delta\theta_{\rm obs, 5}\right)^{-2}T\;,
\end{equation}
where we have plugged in Eq. \eqref{eq:deltatheta} and defined $\Delta\theta_{\rm obs, 5} \equiv\Delta\thetaobs/5$\textdegree.
%which is longer than the orbital period by a factor $R/R_{\rm s}$ where $R_{\rm s}=2GM/c^2$ is the Schwarzschild radius.
Note that for a detectable $\Delta\thetaobs\sim$ few deg, $T_{\rm LT}$ is a factor $\sim 10^2-10^3$ longer than the orbital period $T$ (if $q\gtrsim 1$, the relevant case in our scenario; see Section \ref{sec:estimates} below).
To have $T_{\rm LT}$ $\sim$ few years we need a very close SMBH system with a very short time for gravitational decay (see Eq. \ref{eq:t_GW}) so it is very unlikely to observe such a tight system. The typical timescale for LT precession is indeed longer, $\sim$ 500 years \citep{Begelman++80}. Note that any periodicity due to the LT effect would be modulated on a shorter timescale by the imprint of the orbital speed. If we happen to have a 500 years precession caused by LT, this would be modulated on a few years timescale by moderate angle oscillations of few degrees, and if we happen to have a few years precession due to LT (which is unlikely) this would have strong modulation of several degrees on the months or days timescale.

Hence, the general relativistic effects discussed in this section are all negligible on the few years timescale. This has a simple heuristic explanation: general relativity will always produce effects proportional to $(v/c)^2$, while the dominant effect discussed above is proportional to $(v/c)$. We thus neglect general relativistic effects in the rest of the paper.

\subsection{The geometry of the jet in space and consequences for the periodicity} \label{sec:geojet}

\begin{figure}
\centering
\includegraphics[width=0.5\textwidth]{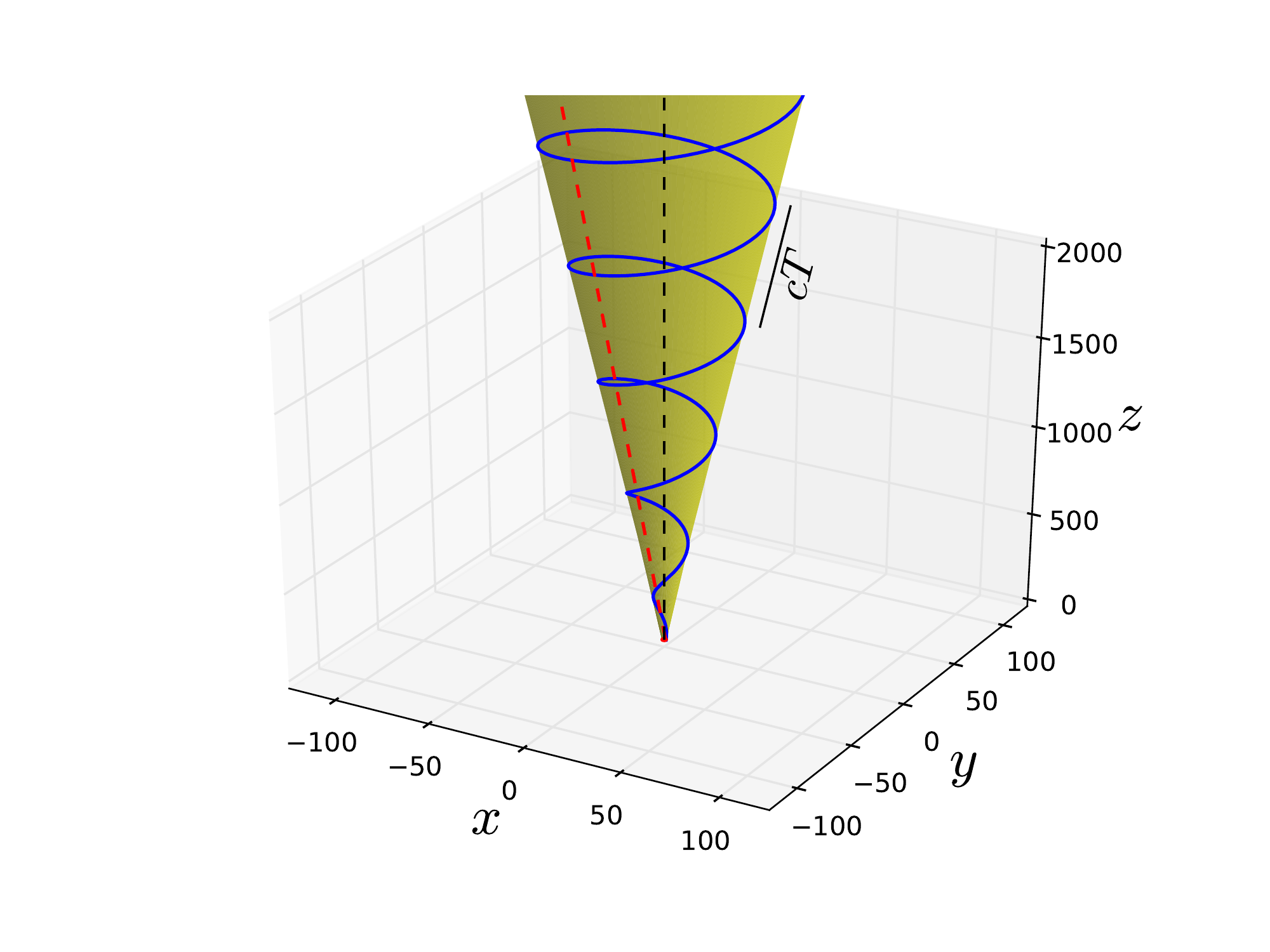}
\caption{Schematic representation of the corkscrew pattern through space formed by the jet. Single particle trajectories are straight lines such as the red dashed line. The launching direction changes with time and the red dashed line sweeps around on the surface of the yellow cone. The result is that the instantaneous shape formed by the jet is the blue solid line. The black-dashed line is the axis of the cone. Note that units on different axes are not to scale (the real pattern is more stretched along the $z$ direction).}
\label{fig:spiral}
\end{figure}

To a first approximation a single emitted particle follows a straight line,\footnote{Apart from general-relativistic effects that are here neglected for the reasons discussed in Section \ref{sec:GR} and deflections due to the gravitational potential of the host galaxy which we also assume can be neglected.} but the launching point moves as the black hole moves on its circular orbit. The consequence is that the jet instantaneous shape is that of a corkscrew through space, see Fig. \ref{fig:spiral}. The radius of the coils increases linearly with the distance (remember that the cone half-opening angle is $v/c$), while the spacing between neighboring coils is $cT$. However one must not confuse the blue helix in Fig. \ref{fig:spiral} with the trajectory followed by a single particle: the latter is a straight line on the surface of a cone (red dashed line in Fig. \ref{fig:spiral}).\footnote{We do not consider here the effect of magnetic fields on the trajectory of individual plasma elements, which may be forced to move on curved paths \citep{CamenzindKrockenberger1992}. Thus our scenario differs from models with single/multiple emitting blobs following helical trajectories \cite[see for example][]{VillataRaiteri1999, Rieger04}. This is not in contrast with the presence of synchrotron emission, which requires magnetic fields: in order to deviate the ejected plasma from straight lines, the energetic of the jet need to be dominated by the Poynting flux, and this is unlikely to be the case for the region where most of the observed beamed radiation comes from \citep[see for example][]{Sikora+2005}.
}
Note also that the actual observed pattern will be distorted since light coming from different points along the spiral takes different times to reach the observer.

Introducing the quantity $T_2\equiv T/2\text{ yr}$, the distance between neighboring coils can be written as
\begin{equation}
cT=1.9\times 10^{18}T_2\text{ cm}\;,
\end{equation}
which for characteristic periods of a few years is of order of $1 \pc$. 

If different wavelengths are produced in regions with different spatial extension, some of them may show periodic behavior while other may not.
If the emitting region extends on scales $\ll cT$ all the points of the emission pattern are observed at the same viewing angle (which is oscillating periodically on the timescale $T$). Thus the light curve can show the same periodicity of the orbital motion. On the other hand, if the emitting region extends on scales $\gtrsim cT$, we are observing many viewing angles at once, and the periodicity is washed out. In this case, however, a well-resolved source might show a characteristic spiral pattern. This idea has been introduced to explain the observed wiggles of jets in radiogalaxies \cite[e.g.][]{BlandfordIcke1978, LuptonGott1982, KaastraRoos1992}.

\subsection{Predicted light curves}

In this subsection we discuss the light curves predicted by our model. We consider an emission pattern extending on scales $\ll cT$ from the base of the jet, so that the viewing angle can be considered approximately the same over all the emission region. 

\subsubsection{$\thetaobs$ as a function of time}

To find the predicted light curves we need to find how $\thetaobs$ as a function of time in our scenario. By choosing the $z$ axis along the orbital angular momentum, the velocity of the ejecta can be written as
\begin{equation}
\pmb{\beta}=\beta\left(\sin\Delta\alpha \cos\Omega t, \sin\Delta\alpha \sin\Omega t,\cos\Delta\alpha\right)\;,
\end{equation}
where $\Omega\equiv 2\pi/T$ is the orbital angular velocity and the half-opening angle of the cone $\Delta\alpha$ is defined in Eq. \eqref{eq:alpha} (see also Fig. \ref{fig:motion}). The velocity of the ejecta gives the direction of the jet axis. Writing the unit vector in the direction of the observer as
\begin{equation}
\hatn=\left(\sin\theta_0,0,\cos\theta_0\right)
\end{equation}
then the cosine of the viewing angle is given by
\begin{equation} \label{eq:costhetaobs}
\cos\thetaobs=\frac{\hatn\cdot\pmb{\beta}}{\beta}=\sin\theta_0\sin\Delta\alpha \cos\Omega t+\cos\theta_0\cos\Delta\alpha\;.
\end{equation}

\subsubsection{Light curves}

Light curves simply follow Eq. \eqref{eq:FB1} where $\thetaobs$ as a function of time is given by \eqref{eq:costhetaobs}. In this section we discuss light curves normalized to their time-averaged value $\bar{F}$, so that overall scaling factors are not important. 

\begin{figure}
\centering
\includegraphics[width=0.5\textwidth]{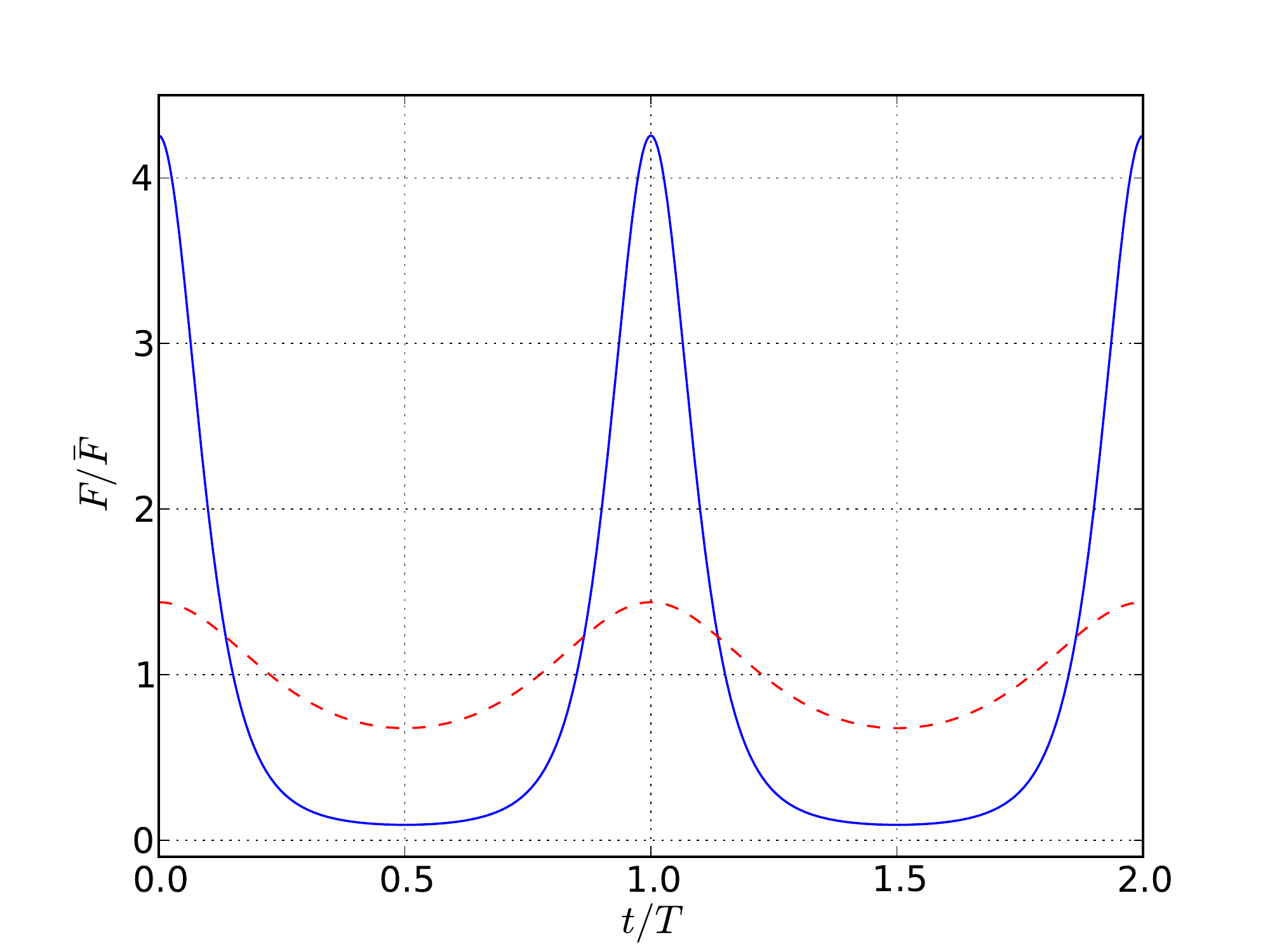}
\caption{Normalized light curves calculated for $\theta_0=\Delta\alpha=2.5\text{\textdegree}$. Different curves correspond to different jet models: $\Gamma_1=20$, $\Gamma_2=2$, $\lambda=0.1$ (blue, solid) and $\Gamma_1=7$, $\Gamma_2=2$, $\lambda=0.1$ (red, dashed). Time is expressed in units of the orbital period $T$.}
\label{fig:lightcurve}
\end{figure}

Let us consider first the total flux over all wavelengths. This follows the factor between square parentheses in Eq. \eqref{eq:FB2}. In Fig. \ref{fig:lightcurve} we show some examples of such light curves calculated using $\theta_0=\Delta\alpha=2.5\text{\textdegree}$ (in this case $\thetaobs$ oscillates between $0\text{\textdegree}$ and $5\text{\textdegree}$). The blue, solid line shows our fiducial model with $\Gamma_1=20$, $\Gamma_2=2$, $\lambda=0.1$. The light curve has an extremely weak dependence on the values of $\Gamma_2$ and $\lambda$, provided that the emissivity at $\thetaobs=0\text{\textdegree}$ is still dominated by the spine. Instead, decreasing (increasing) $\Gamma_1$ decreases (increases) the luminosity contrast $F_{\rm max}/F_{\rm min}$ and increases (decreases) the width of the maxima.\footnote{The maximum and minimum fluxes, $F_{\rm max}$ and $F_{\rm min}$, are obtained for $\thetaobs=0\text{\textdegree}$ and $\thetaobs=5\text{\textdegree}$ respectively.} It is simple to figure out the reason why this happens: while $\thetaobs$ is oscillating, the total flux remains comparable with $F_{\rm max}$ if $\thetaobs\lesssim 1/\Gamma_1$, and hence for a fraction $\sim\pi/\Gamma_1$ of the time. This effect can be clearly seen in Fig. \ref{fig:lightcurve} by comparing the blue solid curve, which is for $\Gamma_1=20$, with the red dashed curve, which is for $\Gamma_1=7$.

Since $\Gamma_1$ controls the luminosity contrast $F_{\rm max}/F_{\rm min}$, one might be led to think that the observation of a periodic light curve allows us to determine the Lorentz factor of the spine, $\Gamma_1$. However, this is not the case since $\Gamma_1$ is degenerate with the geometrical parameters $\theta_0$ and $\Delta\alpha$. For example, the light curve calculated for $\theta_0=2.5\text{\textdegree}$, $\Delta\alpha=2.5\text{\textdegree}$, $\Gamma_1=7$, $\Gamma_2=2$, $\lambda=0.1$ is almost identical to that for $\theta_0=0.8\text{\textdegree}$, $\Delta\alpha=0.8\text{\textdegree}$, $\Gamma_1=20$, $\Gamma_2=2$, $\lambda=0.1$.

Light curves for particular spectral bands can be calculated in a similar way from Eq. \eqref{eq:FB1}. Note that it is often not necessary to know in detail the form of $j_0$, but is enough to know some characteristics of the emission (for example, the index $\eta$ if $j_0$ is a power law, see Eq. \ref{eq:FB3}). We will see an example of this when we calculate the light curves for PG 1553+113 in Section \ref{sec:application} below. 

\subsection{Masses and separation as a function of observables}
\label{sec:estimates}

The entire history of human affairs has a finite extension in time, and even shorter is the history of astronomy. This puts a limit on the longest period that we have been able to observe with our telescopes to date, which is $T\lesssim \text{few yr}$. Moreover, if the model described is to be applied to a real system, we need oscillations of the viewing angle of order $\Delta\thetaobs\sim\text{few deg}$ in order to have easily detectable observational consequences (see Fig. \ref{fig:predictedflux}).

These two facts can be used to constrain mass $M$ and separation $R$ of a binary system to which our model applies. From Eq. \eqref{eq:period} and \eqref{eq:deltatheta} we find for the separation between the SMBH:
\begin{equation}
R=1.3\times 10^{16}\left(\frac{1+q}{q}\right)T_2\left(\Delta\theta_{\rm obs,5}\right)\text{ cm}
\label{eq:binaryradius}
\end{equation}
and the total mass
\begin{equation}
M=1.7\times 10^8\left(\frac{1+q}{q}\right)^3T_2\left(\Delta\theta_{\rm obs,5}\right)^3 M_\odot\;,
\label{eq:binarymass}
\end{equation}
where we have introduced the quantities $\Delta\theta_{\rm obs,5}\equiv\Delta\thetaobs/5\text{\textdegree}$ and $T_2\equiv T/2\text{ yr}$. The numerical factors on the right sides of Eqs. \eqref{eq:binaryradius} and \eqref{eq:binarymass} provide typical values for systems to which our model could apply. 

Finally, a binary system such as the one considered here can decay due to emission of gravitational waves. This also constraints the allowed parameters for our scenario. Assuming circular orbits, the binary SMBH gravitational-wave emission decay time is \citep{PetersMathews63}:
\begin{align}
T_{\rm GW} & = \frac{5}{256}\frac{c^5}{G^3}\frac{R^4}{M_{\rm A}M_{\rm B}M}=\nonumber\\
& = 3.9\times 10^4q\left(\frac{q}{1+q}\right)^3T_2\left(\Delta\theta_{\rm obs,5}\right)^{-5}\text{ yr}\;,
\label{eq:t_GW}
\end{align}
where we have used Eqs. \eqref{eq:binaryradius} and \eqref{eq:binarymass}. In order to avoid extremely short time scales for the orbital decay, we need to assume that the jet is carried by the secondary SMBH (i.e. $q\gtrsim 1$).
This was suggested in the case of OJ 287 \citep{Neronov:2011lr}, the first blazar with clear evidence of periodic variability and a candidate to harbor a system of binary SMBHs.

We can ask whether the few years timescale over which we expect to observe our scenario has a physical meaning or whether it is a mere consequence of selection effects. As we have just seen, shorter periods are unlikely due to orbital decay by gravitational waves emission. Longer periods, even though we would have not been able to observe them yet, are also not allowed in our scenario: the deflection angle depends on the speed of the SMBH and hence for longer periods it becomes too small and therefore undetectable,
%(see Eqs. \ref{eq:alpha} and \ref{eq:v_SMBH})
unless the total mass $M$ is extremely large (to get $T\sim 10^3\text{ yr}$ with $\Delta\thetaobs\sim$ few deg, one needs $M\sim 10^{11}M_\odot$; see Eq. \ref{eq:binarymass}). However there are a number of other effects that can cause precession on longer periods, such as Lense-Thirring \citep[see][and also Section \ref{sec:GR}; timescales of $10^2-10^3$ years]{Begelman++80} or spin-stellar disc interaction \citep[][timescales of $10^7-10^{10}$ years]{MerrittVasiliev2012}. It would be difficult to disentangle these effects observationally (apart from arguments based on the expected timescale). However note also that as we noted in Section \ref{sec:GR} if a periodicity due to LT is present on the $10^2-10^3\text{ years}$ timescale, then it must be accompanied by a few years periodicity due to the orbital motion as depicted in our scenario.

\section{Application to PG 1553+113} \label{sec:application}

\begin{table*}
\caption{Parameters of our representative model for PG 1553+113.}
\begin{tabular} {c c c c c c c c c}
\toprule
$\Gamma_1$ & $\Gamma_2$ & $\lambda$ & $\theta_0$ & $\Delta \alpha$ & $\Omega_{\rm obs}$ & $b$ & $\nu_{\rm P}$ & $\nu_{\rm P} F \left( \nu_{\rm P} \right)$ \\
\midrule
7.0 & 1.1 & 0.1 & 4$\degree$ & 3$\degree$ & 2.88$\,\rm yr^{-1}$ & 0.16 & $7.1\times 10^{14}\text{ Hz}$ & $4.3\times 10^{-12}\text{ erg cm$^{-2}$ s$^{-1}$}$ \\
\bottomrule
\end{tabular} \label{table:1}
\end{table*}

In this section we apply our scenario to the BL-Lac object PG 1553+113. This source has recently been found to show a quasi-periodic modulation ($T_{\rm obs}=2.18\pm 0.08\text{ yr}$) of the light curve at optical-UV and gamma wavelengths \citep{Ackermann++2015}.\footnote{Note that the observed period, $T_{\rm obs}$, is longer by a factor $1+z$ than the orbital period, $T$, where $z$ is the redshift of the source. Consequences of this issue on mass and distance estimates will be examined in Section \ref{sec:param}.} In the radio there are some claims of periodicity (although not in phase with the other frequencies), but the situation is more controversial.

The spectral energy distribution (SED) of PG 1553+113 (at a fixed observation time) is well described by the usual one-zone self-synchro-Compton model (SSC; \citealt{Aleksic++2010, Aleksic++2012, Aleksic++2015}). However, these authors do not consider the quasi-periodic modulation of the light curve on a few years time scales.

Our model is constructed by combining the jet model in Section \ref{sec:jetmodel} and the scenario in Section \ref{sec:binary}. We manually explored the parameter space, and in Table \ref{table:1} we list parameters of a representative model which reproduces both the light curve and the spectra simultaneously reasonably well. This is of course only a by-eye fitting, but is enough for the purposes of showing an example of our scenario in action. 

The parameter space exploration went through the following steps:
(i) we use the same observed frequency, $\Omega_{\rm obs}\equiv 2\pi/T_{\rm obs}$, of \citet{Ackermann++2015};
(ii) $b$, $\nu_{\rm P}$ and $\nu_{\rm P} F \left( \nu_{\rm P} \right)$ are chosen within $30$\% the values obtained with a log-parabolic fit for the most powerful spectrum. In particular, $b$ is approximately constant through different spectra and barely affects other characteristics of the spectra and light curve, so it is easily found by doing a simple log-parabolic fit to one of the spectra.
(iii) we explore the other parameters in the ranges $1\leq\Gamma_2\leq\Gamma_1\leq 30$, $0\leq\lambda\leq 1$, $0\degree\leq \theta_0$, $\Delta\alpha\leq 10\degree$.\footnote{The limits on $\Gamma$ are consistent with the Lorentz factor usually measured in blazars \citep[see for example][]{UrryPadovani1995}. Those on the angles satisfy the following requirements: (i) the blazar is observed at an angle $\lesssim 1/\Gamma$; (ii) the oscillation of the viewing angle does not imply unlikely high masses (see Eq. \ref{eq:binarymass}).}

\begin{figure*}
\centering
\includegraphics[width=\textwidth]{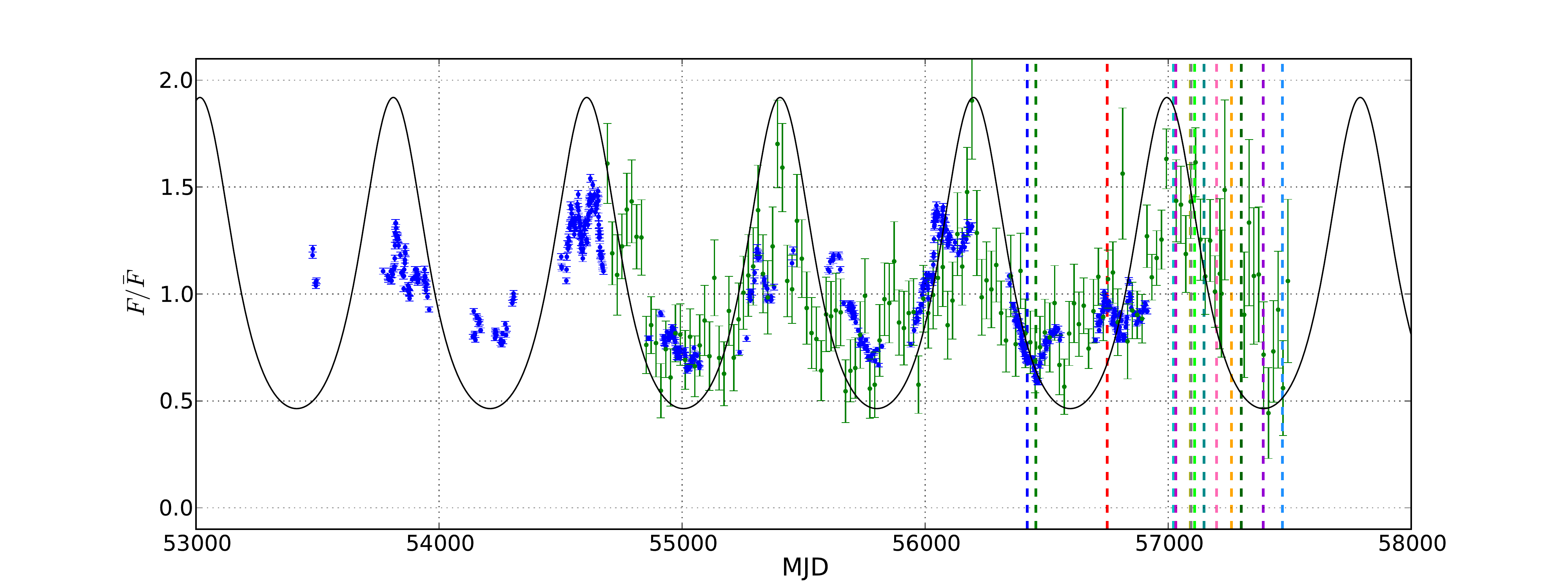}
\caption{Normalized light curve in the optical (blue dots) and gamma (green dots) bands. The vertical lines (and their colours) correspond to the spectra shown in Fig. \ref{fig:blazar_fit}. The black line shows the prediction of our representative model, which also simultaneously reproduces the spectra using the same set of parameters.
Optical points are R-band observations from the Tuorla blazar monitoring program \url{http://users.utu.fi/kani/1m} \citep[extracted from][]{Ackermann++2015}. Gamma-ray data from the {\em Fermi}/LAT observatory at $E>100 \text{ MeV}$ with 20-days bins, extracted from \citet{Ackermann++2015}. From MJD 57219 to 57613 the gamma-ray flux was computed through the ASDC online scientific data analysis \url{http://www.asdc.asi.it}.
}
\label{fig:blazar_LC}
\end{figure*}

\begin{figure*}
\centering
\includegraphics[width=0.49\textwidth]{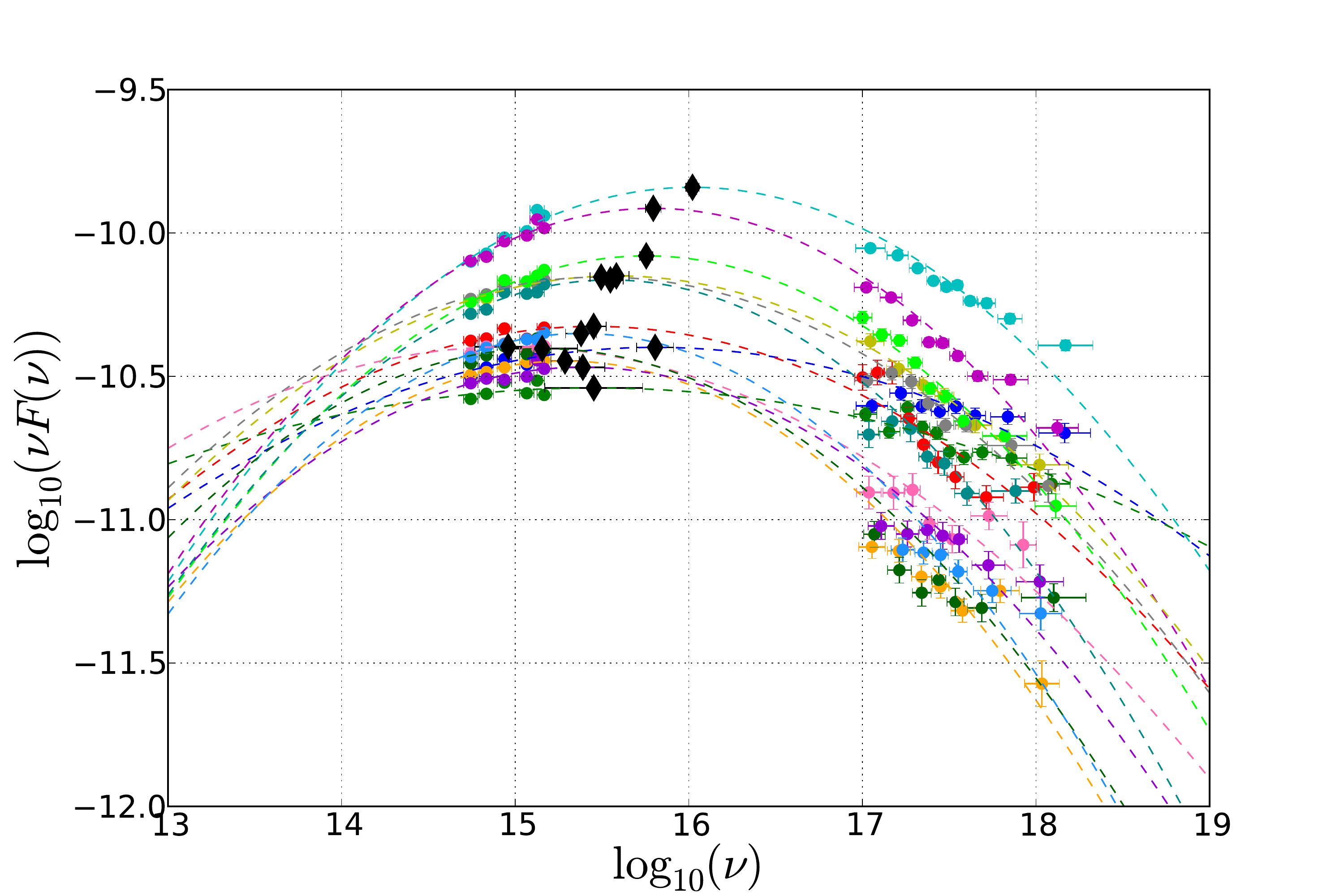}
\includegraphics[width=0.49\textwidth]{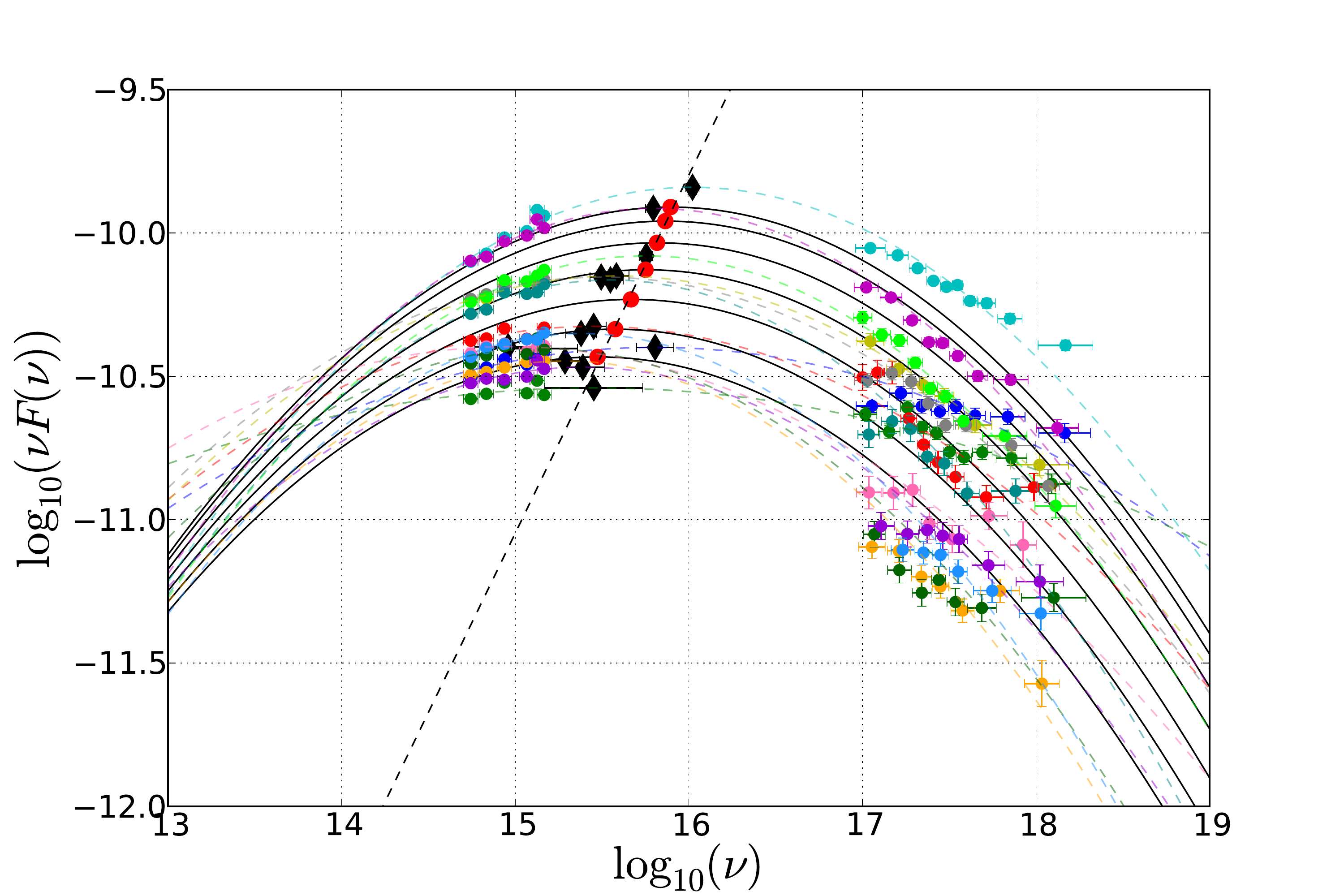}
\caption{Left panel: observed spectra in the optical-UV and X-ray bands, with different colours corresponding to different observation times (see the vertical lines in Fig. \ref{fig:blazar_LC}). The dashed curves show the result of a log-parabolic fit (see Eq. \ref{eq:sed}), whose maxima are marked with black diamonds. Right panel: full black curves show the predictions of our representative model, superimposed on the same spectra shown on the left. The peak of each full black curve is marked with a red dot. The dashed line joins the red dots to help visualisation.}
\label{fig:blazar_fit}
\end{figure*}

\subsection{Light curves}

We start by considering the observed light curves in the optical and gamma ($>100\text{ MeV}$) bands, shown in Fig. \ref{fig:blazar_LC} with blue and green dots respectively. Each light curve is normalized to the time-averaged value of the flux. The quasi-periodic oscillations of the two light curves are in phase with each other. This strengthens the case for a geometrical interpretation for the periodicity such as the one we are proposing. 

The fluxes predicted by our representative model for the optical and gamma bands can be found by integrating Eq. \eqref{eq:FB1} over the appropriate frequency range. However, it turns out that in our case we can approximate both the optical and gamma ray normalized light curves using the expression in square brackets in Eq. \eqref{eq:FB2}.  For optical emission the validity of this approximation can be verified directly by integrating Eq. \eqref{eq:F3}, since it is part of the synchrotron emission for PG 1553+113. For the gamma band, the approximation can be justified from the following consideration. If we look at the gamma spectrum \citep[e.g.][their Fig. 6]{Aleksic++2015}, we see that it has a peak that is always contained within the integration region of Eq. \eqref{eq:FB1}. Hence, most of the value of the integral comes from the region around the peak and extreme of integrations do not really matter.

Thus, the same normalized light curve can reproduce both the optical and gamma light curves. This is nicely verified in Fig. \ref{fig:blazar_LC}, where the blue and green dots follow approximately the same curve. The black line in Fig. \ref{fig:blazar_LC} shows the light curve based on Eq. \eqref{eq:FB2}, where the angle $\thetaobs$ varies in time as predicted by Eq. \eqref{eq:costhetaobs}. The predictions resemble the observations reasonably well, given the uncertainties and the intrinsic optical variability typical of AGNs and blazars. Note that the same model (same parameters) is also able to simultaneously reproduce the spectra, as shown in Section \ref{sec:spectra}. This is a strong constraint: we verified that a better fit could be easily obtained if we only aimed at reproducing the light curve, regardless of the spectra.

Our choice of parameter make $\thetaobs$ oscillate between $1\text{\textdegree}$ and $7\text{\textdegree}$. The parameters are chosen by visual inspection, and no attempt to a more sophisticated analysis has been made. The predicted light curve is quite sensitive to the value of $\theta_0$ and $\Delta\alpha$: for example, changing one of these angles by $\sim 1\text{\textdegree}$ modifies the amplitude of the light curve by $\sim 50\text{\%}$.

As mentioned above, light curve does not show any clear periodicity in the radio band. If spatial extent of the radio emitting region, which tends to increase at low frequencies \cite[e.g.][]{GhiselliniMaraschi1989}, is $\gtrsim cT$ then the periodicity may be washed out for the reasons we explained at the end of Section \ref{sec:geojet}. However, in the case of PG 1553+113 $cT\sim 0.44\text{ pc}$ while the radio core in the well resolved jet of the FRI galaxy M87 is a factor $\sim 10-20$ smaller \citep{Hada++2011}. Thus, to support this interpretation, the radio core of the BL-Lac object PG 1553+113 should be one order of magnitude larger than what is actually observed in M87.

\subsection{Optical-UV-X ray spectrum} \label{sec:spectra}

Here we apply the same representative model of the previous section to the Optical-UV and X-ray SED region, which for PG 1553+113 includes the peak of the synchrotron emission. The data have been collected by the UVOT and XRT instruments onboard the {\em Swift} satellite; details of the reduction and analysis are reported in Appendix \ref{data}.
In the left panel of Fig. \ref{fig:blazar_fit} we show the observed spectra in the optical-UV bands, with different colours corresponding to different observation times (see the vertical lines in Fig. \ref{fig:blazar_LC}). The dashed curves show the result of a simple log-parabolic fit (see Eq. \ref{eq:sed}), whose peaks are marked with black diamonds. These maxima are displaced on a line with slope $\xi\sim 1$ in the $\log(\nu F) - \log(\nu)$ plane: this cannot be explained with a jet with material all moving at the same Lorentz factor, but it can by a jet with a spine-sheath structure (see the discussion in Section \ref{sec:flux}).

The right panel shows the predictions of our representative model (black lines) superimposed to the observations.\footnote{One may argue that the Optical-UV-X ray spectrum predicted from our phenomenological emissivity is then modified since some photons are Compton-scattered to higher energies and are eventually observed in the gamma part of the spectrum. However, from the spectrum of the source \citep[see for example][]{Aleksic++2010, Aleksic++2012, Aleksic++2015} we have estimated that only a negligible fraction $\sim 10^{-4}$ of the synchrotron photons are scattered to higher energies.} The black curves are equally spaced ($\Delta\thetaobs=1\text{\textdegree}$) between $\thetaobs=1\text{\textdegree}$ and $\thetaobs=7\text{\textdegree}$. The peaks of the black lines are marked with red dots and the dashed line joining the dots helps visualisation. The agreement between theory and model is as good as the error bars on the data allow. Our goal is not to provide a precise fit to each single spectrum (which would require a different set of parameters for each spectrum). Instead, we aim to explain simultaneously all the spectra reasonably well with a single set of parameters, focusing on some global features (e.g. the line of the position of the peaks), and to show that the same parameters also reproduce the light curves.

\subsection{Gamma ray spectrum}

A proper analysis of the gamma portion of the spectrum within our model would require the following: (i) to model the $\gamma$-ray emission properly one has to account for the attenuation of $\sim\text{ TeV}$ photons by pair production with the extragalactic background light \cite[see for example][]{Aharonian04}; (ii) the relative contribution to the $\gamma$-ray emission of SSC \citep{Jones++74} and comptonization of the external radiation field from the disk \citep{DermerSchlickeiser93}, the broad line region \citep{Sikora++94} or the dusty torus \citep{Blazejowski2000,Arbeiter2002} is still unclear; (iii) even in the simplest SSC scenario one has to consider re-processing of synchrotron photons from the spine (sheath) by the electron population in the sheath (spine), which depends on some weakly constrained parameters \citep{Sikora++2016}. Investigation of these issues is out of the scope of this work, and hopefully will be the subject of further study.

\subsection{Estimates of system parameters}
\label{sec:param}

From our model we can derive the physical parameters of the system.\footnote{The results presented here have to be interpreted as reasonable estimates of the system parameters, not as a precise measurement. To do this, one would need to explore systematically the parameters degeneracy and to go beyond our simple, spin-aligned scenario, which is outside the scope of this work.} One has to pay attention to observed period $T_{\rm obs}$ is longer than the binary period $T$ (as given by Eq. \ref{eq:period}) by a factor $1+z$, where $z$ is the redshift of the source.
The redshift of PG 1553+113 is not known, but a sound range around $z\simeq 0.5$ was established with different methods \citep{YangWang2010,Danforth++2010,Abramowski++2015,Aliu++2015}. Hence, according to our model, the intrinsic period of the binary is $T\simeq 1.45\text{ yr}$.
Using $T_2=0.73$ and $\Delta\theta_{\rm obs,5}=1.2$ (remember that $\Delta\thetaobs=2\Delta\alpha=6\degree$) in Eq. \eqref{eq:binaryradius} and \eqref{eq:binarymass} we find for the separation and the total mass of the SMBH binary:
\begin{align}
R=1.1\times 10^{16}\left(\frac{1+q}{q}\right)\text{ cm}\;,\\
M=2.1\times 10^8\left(\frac{1+q}{q}\right)^3 M_\odot\;.
\end{align}

The time scale for coalescence due to the emission of gravitational waves can be derived from Eq. \eqref{eq:t_GW}:
\begin{equation}
T_{\rm GW}=1.1\times 10^4q\left(\frac{q}{1+q}\right)^3\text{ yr}\;.
\end{equation}
We stress again that the jet needs to be carried by the secondary SMBH (i.e. $q\gtrsim 1$) in order to avoid an extremely short $T_{\rm GW}$.

\section{Conclusions} \label{sec:conclusion}

In this paper, we have discussed a scenario in which a binary system of SMBHs can give rise to a jet precessing periodically with timescale of $\sim$ few years. The jet is preferably carried by the least massive of the couple and we have discussed several effects that can make it precess. It turns out that the dominant effect (i.e., the one that causes the jet angle to vary by the greatest amount) is simply the imprint of the SMBH orbital speed on the jet. Gravitational deflection and Lense-Thirring precession are second-order effects.

Our scenario applies to systems whose periods are of order of a few years and whose masses are of order of $10^8 M_\odot$. The jet forms a corkscrew pattern in space, and depending on where and on what scales the emission is formed along this pattern we expect a periodic or non-periodic signal. However, when the signal is non-periodic, we still expect some signature of the corkscrew pattern if the source can be resolved.

To make our scenario more concrete, we have modelled the jet with a spine-sheath structure and we have modelled phenomenologically the synchrotron  emission. One of the main features of this modelization is that it allows the peak of the synchrotron spectrum to move not necessarily with a slope of $\xi=3$ in the $\log(\nu F) - \log(\nu)$ plane. Since the binary system scenario and the jet model are independent, in future application it will be possible to use either of these and combine it with a different version of the other, for example involving a different jet structure or a different mechanism for jet precession. 

The observational counterpart of our scenario may be a blazar whose emission varies periodically with time. As an example, we have applied our scenario to the recently-claimed periodic blazar PG 1553+113. We have found that our model can simultaneously explain the light curves in the optical and gamma band and the synchrotron optical-UV spectra at different times. We have also speculated on why the radio emission does not show a clear periodicity, blaming the fact that radio emission is created further out and over longer scales than the optical and gamma emission. We have not attempted to model the gamma spectrum, as this requires an analysis which is outside of the scope of the present paper. Finally, we have given estimates of the masses and separation for the system.

We have not discussed the mechanism that produces the jet. Instead, we merely assumed a jet is present. Such assumption is justified by the fact that we do see jets occurring in nature, but the mechanism that produces such jets is still a major open problem in astrophysics \cite[see for example][]{Livio2009}. In a binary system scenario such as the one discussed in this paper, the situation is even more complicated and less well studied. Hopefully, some much needed progress will be made in this direction over the next years. 

Other directions for future work include (i) the study of more sophisticate emission models, including a self-consistent scenario for the emissivity of gamma photons; (ii) a more careful consideration of disk and jet dynamics/stability in binary systems \cite[see for example][]{Farris++2011, Noble++2012}; (iii) the application of our model to other systems. Up to date the other most spectacular example of periodic blazar is OJ 287 ($T\sim 12\text{ yr}$). However, due to the presence of a narrow double peak at maximum brightness, the periodicity is more commonly interpreted as a result of the smaller SMBH punching the accretion disk of the companion during an eccentric orbit \cite[see for example][]{Valtonen++2006,Valtonen++2008}. Hopefully, more detections of similar objects \cite[see for example][]{Sandrinelli++2016} will improve our understanding of the underlying physical mechanisms that cause observed periodicities.

%% file: images/Jet.pdf_tex
%% Creator: Inkscape inkscape 0.91, www.inkscape.org
%% Accompanies image file 'Jet.pdf' (pdf, eps, ps)
\begingroup%
  \makeatletter%
  \providecommand\color[2][]{%
    \errmessage{(Inkscape) Color is used for the text in Inkscape, but the package 'color.sty' is not loaded}%
    \renewcommand\color[2][]{}%
  }%
  \providecommand\transparent[1]{%
    \errmessage{(Inkscape) Transparency is used (non-zero) for the text in Inkscape, but the package 'transparent.sty' is not loaded}%
    \renewcommand\transparent[1]{}%
  }%
  \providecommand\rotatebox[2]{#2}%
  \ifx\svgwidth\undefined%
    \setlength{\unitlength}{344.413493bp}%
    \ifx\svgscale\undefined%
      \relax%
    \else%
      \setlength{\unitlength}{\unitlength * \real{\svgscale}}%
    \fi%
  \else%
    \setlength{\unitlength}{\svgwidth}%
  \fi%
  \global\let\svgwidth\undefined%
  \global\let\svgscale\undefined%
  \makeatother%
  \begin{picture}(1,0.58540297)%
    \put(-1.05584028,0.51144154){\color[rgb]{0,0,0}\makebox(0,0)[lt]{\begin{minipage}{0.05578437\unitlength}\raggedright \end{minipage}}}%
    \put(0,0){\includegraphics[width=\unitlength,page=1]{Jet.pdf}}%
    \put(0.88895409,0.37224271){\color[rgb]{0,0,0}\makebox(0,0)[lt]{\begin{minipage}{0.03430041\unitlength}\raggedright \end{minipage}}}%
    \put(0.8463307,0.35354609){\color[rgb]{0,0,0}\makebox(0,0)[lb]{\smash{$\Gamma_1$}}}%
    \put(0.82310281,0.4603944){\color[rgb]{0,0,0}\makebox(0,0)[lb]{\smash{$\Gamma_2$}}}%
    \put(0.82310281,0.2420522){\color[rgb]{0,0,0}\makebox(0,0)[lb]{\smash{$\Gamma_2$}}}%
    \put(0.14463459,0.56775522){\color[rgb]{0,0,0}\makebox(0,0)[lb]{\smash{\Large{$\Sigma$}}}}%
    \put(0,0){\includegraphics[width=\unitlength,page=2]{Jet.pdf}}%
    \put(0.08107322,0.35834613){\color[rgb]{0,0,0}\makebox(0,0)[lb]{\smash{$R_1$}}}%
    \put(0.01600488,0.33484164){\color[rgb]{0,0,0}\makebox(0,0)[lb]{\smash{$R_2$}}}%
    \put(0,0){\includegraphics[width=\unitlength,page=3]{Jet.pdf}}%
    \put(0.38768145,0.29330803){\color[rgb]{0,0,0}\makebox(0,0)[lb]{\smash{$\theta_{\rm obs}$}}}%
    \put(0.50784182,0.15153256){\color[rgb]{0,0,0}\rotatebox{-46.5089934}{\makebox(0,0)[lb]{\smash{To observer }}}}%
  \end{picture}%
\endgroup%

%% file: images/motion.pdf_tex
%% Creator: Inkscape inkscape 0.91, www.inkscape.org
%% Accompanies image file 'motion.pdf' (pdf, eps, ps)
\begingroup%
  \makeatletter%
  \providecommand\color[2][]{%
    \errmessage{(Inkscape) Color is used for the text in Inkscape, but the package 'color.sty' is not loaded}%
    \renewcommand\color[2][]{}%
  }%
  \providecommand\transparent[1]{%
    \errmessage{(Inkscape) Transparency is used (non-zero) for the text in Inkscape, but the package 'transparent.sty' is not loaded}%
    \renewcommand\transparent[1]{}%
  }%
  \providecommand\rotatebox[2]{#2}%
  \ifx\svgwidth\undefined%
    \setlength{\unitlength}{244.86987737bp}%
    \ifx\svgscale\undefined%
      \relax%
    \else%
      \setlength{\unitlength}{\unitlength * \real{\svgscale}}%
    \fi%
  \else%
    \setlength{\unitlength}{\svgwidth}%
  \fi%
  \global\let\svgwidth\undefined%
  \global\let\svgscale\undefined%
  \makeatother%
  \begin{picture}(1,1.54073341)%
    \put(-1.81391171,-0.28466336){\color[rgb]{0,0,0}\makebox(0,0)[lt]{\begin{minipage}{0.07846163\unitlength}\raggedright \end{minipage}}}%
    \put(0.92147342,-0.4804488){\color[rgb]{0,0,0}\makebox(0,0)[lt]{\begin{minipage}{0.04824409\unitlength}\raggedright \end{minipage}}}%
    \put(0.94961439,-0.89305279){\color[rgb]{0,0,0}\makebox(0,0)[lb]{\smash{}}}%
    \put(0,0){\includegraphics[width=\unitlength,page=1]{motion.pdf}}%
    \put(0.29185732,0.08985864){\color[rgb]{0,0,0}\makebox(0,0)[lb]{\smash{$v$}}}%
    \put(0.55435374,0.30392415){\color[rgb]{0,0,0}\makebox(0,0)[lb]{\smash{spin}}}%
    \put(0.50654365,0.00679575){\color[rgb]{0,0,0}\makebox(0,0)[lb]{\smash{Jet-carrying black hole}}}%
    \put(0,0){\includegraphics[width=\unitlength,page=2]{motion.pdf}}%
    \put(0.00059024,0.89869802){\color[rgb]{0,0,0}\makebox(0,0)[lb]{\smash{jet direction}}}%
    \put(0,0){\includegraphics[width=\unitlength,page=3]{motion.pdf}}%
    \put(0.44612612,0.34295792){\color[rgb]{0,0,0}\makebox(0,0)[lb]{\smash{$\Delta\alpha$}}}%
    \put(0,0){\includegraphics[width=\unitlength,page=4]{motion.pdf}}%
    \put(0.38551088,1.50283904){\color[rgb]{0,0,0}\makebox(0,0)[lb]{\smash{observer}}}%
    \put(0,0){\includegraphics[width=\unitlength,page=5]{motion.pdf}}%
    \put(0.39694552,0.4622864){\color[rgb]{0,0,0}\makebox(0,0)[lb]{\smash{$\theta_{\rm obs}$}}}%
    \put(0,0){\includegraphics[width=\unitlength,page=6]{motion.pdf}}%
    \put(0.45411874,0.6125703){\color[rgb]{0,0,0}\makebox(0,0)[lb]{\smash{$\theta_0$}}}%
    \put(-0.36590862,0.73835138){\color[rgb]{0,0,0}\makebox(0,0)[lb]{\smash{}}}%
  \end{picture}%
\endgroup%